**Accurate Vertical Nanoelectromechanical Measurements**

R. Proksch,[1,a)] R. Wagner,[2] and J. Lefever[1]

[1]*Asylum Research an Oxford Instruments Company, Santa Barbara, CA, 93117 USA*

[2]*School of Mechanical Engineering, Purdue University, West Lafayette, IN, 47907 USA*

Piezoresponse Force Microscopy (PFM) is capable of detecting strains in these materials, down to the picometer range. With the emergence of weaker materials, the smaller signals associated with them have uncovered ubiquitous crosstalk challenges that limit accuracy of measurements and that can even mask them entirely. Previously, using an interferometric displacement sensor (IDS), we demonstrated the existence of a special spot position immediately above the tip of the cantilever, where the signal due to body-electrostatic (BES) forces is nullified. Placing the IDS detection spot at this location allows sensitive and BES artifact-free electromechanical measurements. We denote this position as $x_{IDS}/L = 1$, where $x_{IDS}$ is the spot position along the cantilever and $L$ is the distance between the base and tip. Recently, a similar approach has been proposed for BES nullification for the more commonly used optical beam deflection (OBD) technique, with a different null position at $x_{OBD}/L \approx 0.6$. In the present study, a large number of automated, sub-resonance spot position dependent measurements were conducted on periodically poled lithium niobate (PPLN). In this work, both IDS and OBD responses were measured simultaneously, allowing direct comparisons of the two approaches. In these extensive measurements, for the IDS, we routinely observed $x_{IDS}/L \approx 1$. In contrast, the OBD null position ranged over a significant fraction of the cantilever length. Worryingly, the magnitudes of the amplitudes measured at the respective null positions were typically different, often by as much as 100%. Theoretically, we explain these results by invoking the presence of both BES and in-plane forces electromechanical forces acting on the tip using an Euler-Bernoulli cantilever beam model. Notably, the IDS measurements support the electromechanical response of lithium niobate predicted with a rigorous electro-elastic model of a sharp PFM tip in the strong indentation contact limit ($d_{eff} \approx 12\,pm/V$, Kalinin et al. Phys. Rev. B 70, 184101 (2004)).

## I. INTRODUCTION

Accurate measurements of the nanoscale electromechanical coupling in piezo and ferroelectrics, twisted 2D layers, quantum and biological materials are of both fundamental scientific and applied importance. More than thirty years ago, Güthner and Dransfield[1] demonstrated piezoresponse force microscopy (PFM) on ferroelectric polymers, marking its emergence as the leading non-destructive technique for exploring electromechanics at the nanoscale. Initially, PFM focused on classical ferroelectric materials with relatively high inverse piezoelectric coefficients ($d_{eff} \geq 10 pm/V$. However, over the past decade, there has been a shift towards investigating weaker materials driven by factors such as computing and memory beyond Moore's law,[2] photonic computing,[3], [4] energy storage and production, bio-electromechanics, MEMS devices (for example, the rapid growth in the number of bulk acoustic resonator filters for 5G and beyond communication) as well as 2D materials including twisted quantum materials.[5]

This trend towards weaker materials and smaller length scales has revealed various measurement artifacts, including instrumental cross-coupling, long-range electrostatic effects, thermal influences, frictional forces, and cantilever dynamics. These artifacts, when mistakenly interpreted as an electromechanical response, can lead to inaccurate conclusions about the nanoscale functionality of a material – sometimes referred to as "strange ferroelectrics." [6] Even when the materials are piezo- or ferro-electric, the common practice of reporting uncalibrated measurements, with the piezoresponse amplitude denoted in "arbitrary units," adds to the confusion when comparing measurements of similar or identical materials.

---

a) Author to whom correspondence should be addressed. Electronic mail: roger.proksch@oxinst.com

In the current work, we show, with a combination of extensive automated experiments and analytical, linear Euler-Bernoulli beam theory, that the combination of vertical, longitudinal, and long-range body electrostatic (BES) forces – as are likely present in most practical PFM measurements – can explain a great deal of the variation researchers have observed over the past three decades of PFM. In particular, the combination of longitudinal (in-plane) forces and BES forces lead to large errors in the OBD measurements while, for IDS measurements with the spot positioned at or near the tip ($x/L \approx 1$) measurements are largely immune to these effects.

PFM is based on the atomic force microscope (AFM), a powerful tool for studying the structure and function of surfaces at the nanoscale. It relies on a sharp tip that can localize interactions with the surface, and the ability to measure cantilever motion from microns down to picometers. The first AFM described by Binnig et al.[7] made use of a tunneling detector to measure cantilever displacement, but this method was complex and subject to limitations. Optical methods have since supplanted the original detector technology and the most common method for measuring cantilever motion is the optical beam deflection (OBD) method,[8, 9] which uses the reflected angle of a laser focused on the back of the cantilever to determine the cantilever tip position. This method measures the angular change of the cantilever and requires assumptions about the cantilever vibration shape to infer the vertical displacement of the tip. Interferometric displacement sensors (IDS), while rare in commercial systems, have the advantages of directly measuring displacement and of having a built-in calibration based on the wavelength of the light source.[10,11, 12, 13, 14]

PFM relies on the converse piezoelectric effect, where the application of a localized electric field induces a mechanical deformation in a material, which is in turn detected by the same tip. The measured signal is related to the local piezoelectric tensor, which is a second-rank tensor that describes the material's response to an applied electric field. The piezoelectric tensor is a fundamental property of piezoelectric materials and is related to the crystal structure of the material. The tensor has three principal axes and contains six independent components that relate the electrical polarization and mechanical strain in the material. In general, this means that, in response to an applied bias, the tip will experience a three-dimensional force. Reported values for these responses are typically measured in bulk materials with large electrodes on the sample. As discussed below, other factors, such as the tip-sample stiffness and voltage drops, along with the sharp tip-sample contact will affect the sensitivity measured by the probe.

Quantitative electromechanical response has been a long-standing goal for the PFM community. They require consideration of the electro-elastic fields generated inside a material due to tip-induced indentation. Kalinin at al.[15] discussed two cases: weak indentation and strong indentation. In the weak indentation case, the electrostatic field distribution is calculated using an image charge model. In the strong indentation case, the coupled electro-elastic problem for piezoelectric indentation is solved to obtain the electric field and strain distribution in the ferroelectric material. Specifically, they predicted $d_{eff} \approx 12pm/V$ for the hard indentation limit sensitivity of $LiNbO_3$ (LNO). While these analytical solutions have contributed to our understanding of PFM contact mechanics and imaging mechanisms, experimental verification of the theoretical predictions have remained elusive, even for the simple case of stoichiometric LNO, where experimental reports range from $2 - 30pm/V$.

The converse piezoelectric effect induced by the electric field of the tip in PFM can lead to both in-plane and out-of-plane surface displacements.[16, 17] Note that much of the literature refers to the in-plane motion leading to "buckling" of the cantilever. Since buckling refers to a different, stability phenomenon for loaded beams, we adopt the term "longitudinal" forces and strains to describe in-plane motion parallel to the long cantilever axis. In other work, the terminology "lateral" is synonymous with "in-plane". In this case, we also differentiate "lateral" motion as in-plane motion perpendicular to the long axis of the cantilever and "longitudinal" as parallel.

The problem of separating the vertical and longitudinal cantilever response in the PFM signal is still unresolved and often ignored, which hinders the development of quantitative piezoresponse measurement. There have been a number of approaches to characterize longitudinal piezoresponse, generally involving two different approaches, the first being referred to as “angle-resolved pfm”, where measurements are made at different cantilever -sample angles. [18], [19], [16], [20], [17], [21], [22] This approach can both provide a vector map of the sample response and serve as a detector for the existence of measurable longitudinal forces. A significant downside is that it requires alignment of the tip and sample at each new angle and, practically, drift and distortion correction of the images.

Another common artifact in electrical mechanical measurements comes from delocalized (long-range) electrostatic forces between the body of the cantilever and the sample surface charge. In addition, the drive frequency of electrical excitation can have a profound effect on the measured signal.[23,24] Since the frequency response of most ferroelectric samples should be flat into the GHz range, features observed in the AFM system transfer function in the kHz or MHz ranges must originate from the microscope itself, most notably the dynamics of the AFM cantilever, instead of from ferroelectric sample properties.[25, 26] To minimize the effects of cantilever dynamics on the ferroelectric signal, single-frequency PFM is commonly limited to operation at a few hundred kHz or lower, [27] with some exceptions. [28] Techniques that track the cantilever’s resonance frequency such as dual AC resonance tracking[29] (DART) and band excitation[30] (BE) reduce the severity of the problem. However, quantitative interpretation of the piezoresponse in these methods requires dynamic modeling involving assumptions about the structure, geometry, boundary conditions, and external forcing of the cantilever. [31], [32], [33]

The PFM community has developed several approaches for minimizing or eliminating long-range electrostatic artifacts [6]. One approach is to use a low frequency bias modulation, which operates at a frequency well below the first contact resonance of the cantilever. Another method is to use smaller cantilevers, which can reduce the electrostatic coupling between the tip and sample. Longer tips can be used to increase the distance between the cantilever body and sample, which reduces the capacitance. Shielded probes are another option, which may reduce the capacitance but are more expensive and not as well developed as conventional cantilevers. Stiffer cantilevers can reduce the effect of long-range electrostatic forces on cantilever motion but may not be suitable for thin films and softer materials as the high loading force may damage the sample. Positioning the OBD (optical beam deflection) spot closer to the base of the cantilever can reduce the effect of nodal lines on measured phase and amplitude, but it comes at the cost of a reduction in sensitivity.[34] Scanning along the edge of a sample can help minimize the long-range electrical effects. Measurements at different sample rotation angles could provide insight by varying the body-charge coupling. Despite much progress, electrostatic artifacts remain a significant challenge in PFM. [35], [26] Misinterpreting the electrostatic, cantilever dynamics and background signals as a tip displacement can lead to incorrect estimation of the piezoelectric sensitivity and relative phase response and the response of the cantilever to in-plane strains, electrostatic forces, background crosstalk and poorly characterized cantilever dynamics can overwhelm the PFM signal of interest.

One of the variables at the AFM experimentalist’s fingertips is the location of the optical spot on the cantilever. In many situations, the default has been to place the spot near the end of the cantilever since this is the position of maximum sensitivity for an OBD detector with a small spot measuring both for an end-loaded, sub-resonant mode and for the first resonant mode of a freely vibrating cantilever. However, there are many situations where the end of the cantilever is not an optimum position. One important example was pointed out by Schäffer and Hansma,[36] where they pointed out the optimum position for the spot was near the middle of the cantilever ($x/L \approx 0.5$) and the spot width was equivalent to the cantilever length. This same analysis was extended to account for higher resonant modes.[37] In particular, the OBD spot position could be used to enhance some resonant mode sensitivities and suppress others.[38] While Killgore et al,[39] coined the term “blind spot”, we follow Huey and refer to these locations as “null points” since they are locations where the detector sensitivity to a particular mode vanishes. Null points exist for surface-coupled cantilevers, as well as freely vibrating levers, as will be discussed below.

Similar to both Huey et al. and Killgore et al., a method for finding the null point where longitudinal contributions to the OBD signal vanish, while still being sensitive to the vertical response was first described by Nath et al. [16] and later by Alikin et al. [20], [17] As discussed above, it involves placement of the OBD spot on the back of the cantilever to a location where the measured response to the longitudinally driven mode is suppressed. Nath et al. studied the piezoresponse of a cantilever due to in-plane motion and noted that although the longitudinal (in their language, "buckling") motion is a result of in-plane strains (in-plane piezo-response), the output signal is strongly coupled in the out-of-plane PFM images. They defined a "fulcrum point" of the longitudinal vibration to be at a point at roughly $x/L \approx 0.6$. When the OBD spot was at this position, they demonstrated that the in-plane response was largely nullified and contained only information on the vertical response of the sample. More recently, Alikin et al.[17] examined the sensitivity of the cantilever for both in-plane and out-of-plane piezoresponse contributions to the vertical PFM response. The PFM mode with OBD laser beam focused close to the cantilever free end was found to mix in-plane and out-of-plane piezoresponse, making the vertical piezoresponse phase dependent on polarization orientation. Theoretically and experimentally, they explored different OBD laser positions to eliminate the longitudinal response and capture the vertical response. Specifically, they suggested positioning the OBD laser spot at $x/L \approx 0.59$, similar to the value described by Nath et al.

In an effort to mitigate the effects of BES forces on OBD PFM measurements, Huey et al. [35] noted that placing the OBD spot at $x/L \approx 0.6$, where $L$ is the distance between the base and the and $x$ is the spot position closer to the base rather than at the tip can reduce or even eliminate the influence of electrostatic artifacts. Recently, Killgore et al. [39] performed an extensive study of this null point (and coined the label "electrostatic blind spot", as mentioned above). They went on to suggest that this positioning, along with the lever sensitivity being calibrated with a force versus distance curve to calibrate the inverse optical lever sensitivity calibration (nm/V), at the null point location allowed quantitative electromechanical measurements. They concluded that OBD measurements could be made to be as quantitative as interferometric measurements, discussed below.

A very different approach to eliminating electrostatic artifacts and improving the reproducibility of electromechanical measurements is a metrological AFM[40] that combines a conventional OBD sensor with a in an Interferometric Displacement Sensor (IDS) based around a laser Doppler vibrometer.[41] A key advantage of interferometric methods is that the sensitivity is intrinsically and accurately calibrated, since the calibration is based on the well-defined wavelength of light. Interferometry measures the tip velocity (or displacement) directly and therefore requires no assumptions about the cantilever mode shape. If the IDS laser spot is directly above the cantilever tip, then tip displacement can be directly measured. Relative to OBD measurements, the IDS is less sensitive to changes in spot size. More importantly, because the IDS measurement is encoded as a frequency (Doppler) shift of the helium-neon laser, the sensitivity is highly accurate and does not change with the optical reflectivity of the cantilever nor with the laser power. While we used a doppler vibrometer, we anticipate that other types of interferometers will have similar benefits and perhaps even better performance in other ways. Since that initial publication, this instrument has successfully been used to explore a wide variety of challenging functional materials systems including studies of fundamental limitations in the detection of nanoscale electromechanical response [42], [6], [43], [44], beyond Moore's law computing, [45], photovoltaics, [46], [47] energy storage and low dimensional ferroelectrics[48], [49].

## II EXPERIMENTAL METHODS

PFM measurements are typically made with the cantilever in contact mode, while the tip-sample voltage is modulated with a periodic tip bias $V_{tip} = V_{DC} + V_{AC}\cos(\omega t)$. On the low frequency side, the modulation frequency is usually chosen to be well above the contact mode feedback bandwidth, typically at least a few kHz. For the microscope and settings in this work, that minimum frequency is ~10kHz. The modulated bias generates an oscillating electric field below the tip which in turn leads to localized deformations in the sample surface. These local oscillations act as a mechanical drive on the tip that, in turn, acts as a mechanical drive for the cantilever. The response of the cantilever to this modulated bias is measured with a detector (as discussed above) and a lock-in amplifier. The cantilever motion at the modulation frequency of the bias has the form $A_{resp} = A_{1\omega}\cos(\omega t + \phi)$. When the response is dominated by piezoelectric strain at the tip, the amplitude $A_{1\omega}$, is proportional to the "effective" vertical converse piezo sensitivity, $d_{z,eff}$ and the periodic(AC) tip bias, $A_{1\omega} = V_{AC}d_{z,eff}$, while the phase $\phi$, of the electromechanical response of the surface is indicative of the sample polarization orientation.

To eliminate as many uncontrolled variables as possible we confined the PFM measurements to one specific sample of PPLN and one particular type of diamond-coated, conductive cantilever. [50] Both samples and probes have the advantage of being commercialized, hopefully allowing easy duplication of the experimental results. The choice of cantilever is well-known to affect the quality of AFM measurements in general and certainly in the particular case of electromechanical measurements. Electromechanical measurements often involve large loads, along with significant in-plane forces. Since this study is meant to be a quantitative comparison as a function of load, spot position and other variables. Because of this requirement, it seemed critical to choose a single type of lever, particularly one that provided extremely good wear characteristics, we chose a doped diamond cantilever with high conductivity and excellent wear characteristics (Adama AD2.8). The periodically poled LNO (PPLN) used here is commercially available and frequently used as a calibration sample. LNO belongs to the point group 3m, which has four independent components in the piezoelectric matrix: $d_{15} \approx 69pm/V$, $d_{22} \approx 21pm/V$, $d_{31} \approx -1pm/V$, and $d_{33} \approx 7pm/V$.[15] The vertical PFM signal measured far away from domain walls will contain contributions from all four independent components of the piezoelectric matrix. Typically, this combined sensitivity is designated as an "effective" [51], [21] out of plane sensitivity, $d_{eff}^{z} \approx 12pm/V$, as estimated by Kalinin et al. for LNO.[15] As was also noted by Jungk et al. in a more qualitative but more easily interpretable statement, since one of the electrodes in the case of PFM is a sharp, conductive tip, the applied electric field in the crystal is highly inhomogeneous,[52] and as a consequence, the piezoresponse is clamped.[53],[16]

In real-world IDS measurements, it is common to measure $d_{eff}^{z} < 12pm/V$ for LNO. This reduction in electromechanical response may be caused by a number of factors. Practical electromechanical measurements are made in imperfect conditions on imperfect surfaces. In this context, the electrode is commonly supplied by the probe tip itself. This experimental setup differs considerably from that of a bulk sample and electrode.

FIG.1 shows a method of testing conductive probes for electromechanical measurements. In FIG. 1(a), the plots are very repeatable and become independent of load above ~100nN. This simple scaling is expected for the hard indentation limit of PFM. These measurements, made with a drive bias, $V_{AC} = 10V$, yielded $d_{eff} \approx 12pm/V$, as expected from the Kalinin et al. limit. In practice, this is not always the case, many measurements show a reduced response, presumably due to the factors discussed in the previous section. FIG. 1 (b) shows the response of a metal coated cantilever that was less robustly manufactured. There is considerable irreproducible variation in the response as a function of the load. In one of the cases, the sensitivity briefly jumps to ~20pm/V. One can speculate as to the origin of this jump, but it was transitory in the force curve and was not repeatable. As an interesting aside, this method can be used as a good test of the quality of conductive cantilevers for other electrical measurements.

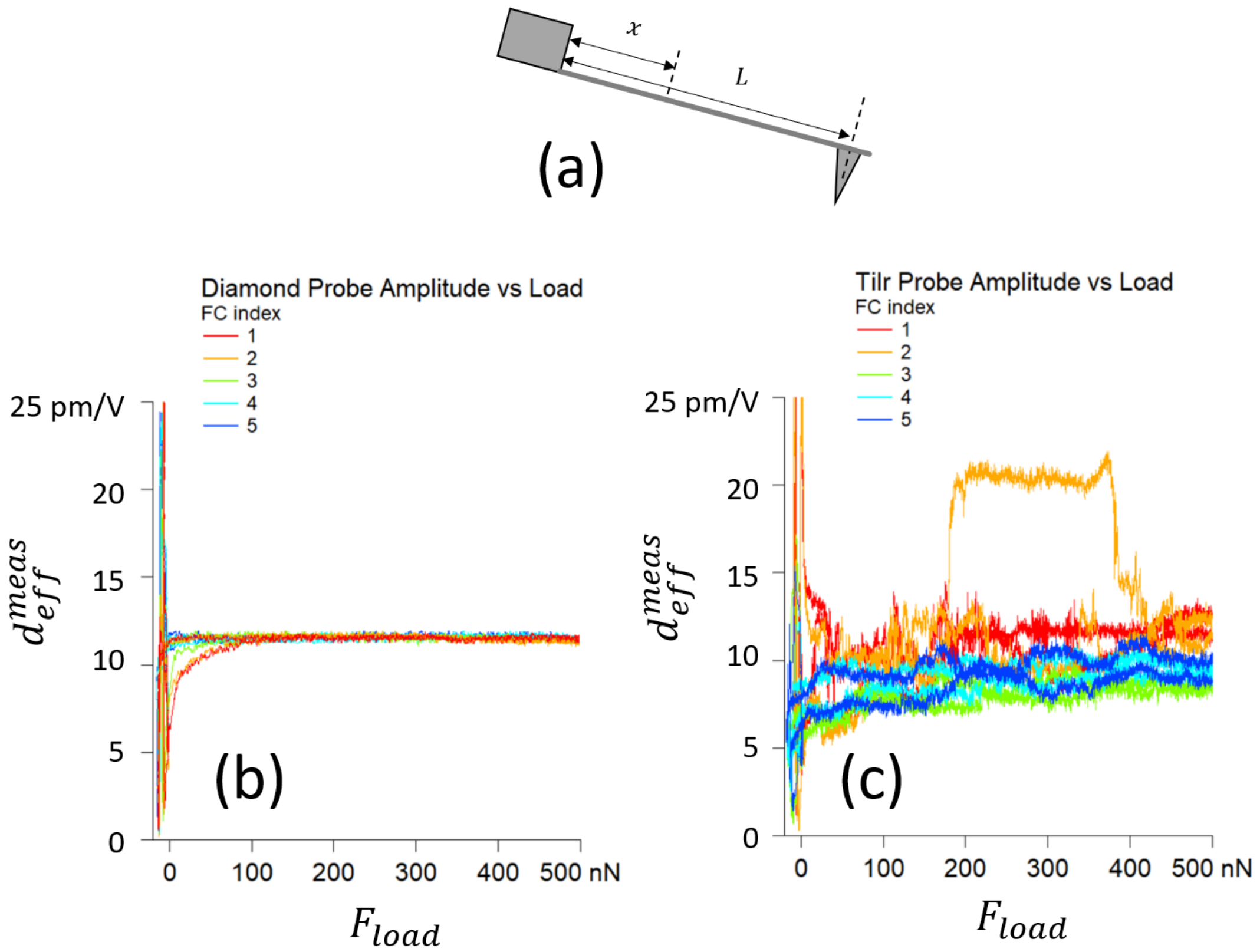

FIG. 1 (a) shows the spot positioning coordinate system. A spot positioned at $x/L = 1$ is over the tip. (b) shows five plots of the PFM IDS amplitude measured as a function of load over PPLN with (c) the nano-diamond probe type used in this study and (b) another popular, TiIr metal alloy coated lever commonly used for conductive and electromechanical measurements.

In practice, the datasets where spot positions were systematically stepped along the length of the required levers that were "seasoned". By seasoning, we mean that the levers had already been used to scan numerous images and were observed to maintain a roughly constant response, as tested with the IDS at $x/L \approx 1$.

As discussed above, we systematically compared PFM measurements on a PPLN sample using IDS and BESNP OBD. We used the three methods proposed by Killgore et al. to find the OBD null point. Specifically, we started with an initial guess of $x_{OBD}/L \approx 0.6$ and $x_{IDS}/L \approx 1$, followed by three different refinement methods defined below.

### A. Method 1 - Equal Amplitude Method (EAM)

The equal amplitude method (EAM) involves adjusting the laser position until oppositely poled domains are equal in amplitude. This assumes prior knowledge that the opposite domains exhibit equal coupling coefficients.

In this method, we first estimated the null point position by finding a location where the domains were unequal and then moving the spot position until the sign of that inequality was reversed. Those two locations became the start and the end positions of a systematic stepping measurement. This approach was an improved version of mode-mapping studies we and others have performed in the past, where the spot-positions of both the OBD and IDS detectors were systematically varied and the spot position dependent responses were measured. [54], [55], [34] These previous approaches, engaged the cantilever on the surface using the OBD detector as the error signal and then disabled the z-feedback loop (feedback-off) while the OBD spot was scanned over different positions. In the case of a zero drift AFM, that is fine. In real

AFMs, there is always some lateral and vertical drift in the tip-sample position as well as drift in the detector. In one study, [34] the detector drift was managed by (i) measuring the before and after free-air deflection voltage and linearly extrapolating between those to estimate the load at the various intermediate, feedback-off positions and (ii) choosing loads large enough that the drift in the detector was deemed insignificant (typically <5% change). This approach limited the practical measurement times and if the drift was too significant over a profile measurement, the data was usually discarded and needed to be repeated.

In the present work, an improved approach was taken that allowed continuous feedback using the OBD sensor. Briefly, after the initialization step, we performed a force curve that allowed calibration of the OBD $InvOLS$ ("inverse optical lever sensitivity", with units of meters/Volt). This value, coupled with the previously measured spring constant of the cantilever allowed the OBD deflection voltage to be estimated as $\Delta V = F_{load}/(k_{cant} \cdot InvOLS)$, where $F_{load}$ is the loading force of the tip against the sample, $k_{cant}$ is the cantilever spring constant. In addition to the experimental drift discussed in the previous paragraph, another shortcoming of the OBD method is that local variations in the reflectivity or roughness of the cantilever can cause variations in the $InvOLS$. Measuring a force curve at each spot position in large part corrects for these variations. Insensitivity to reflectance variations is one advantage of an interferometric approach; variations in the reflectivity of the cantilever translate into variations in the noise level. During this same force curve, the amplitude and phase were also collected as a measure of the tip-sample coupling. Once engaged at a given loading force $F_{load}$, other measurements including surface tunes and images were acquired. Once these measurements were completed, the cantilever was withdrawn from the surface and the spot positions updated. This process was repeated for as many positions along the lever as chosen, typically between 10 and 50 in this study.

The spot positions of the OBD and IDS were typically changed simultaneously since the motorized laser stage on our instrument moves the cantilever relative to both the optical spots the same distance. This is convenient for observing relative sensitivities to spot position. Unlike in proceeding measurements, the OBD and IDS measurements were made simultaneously through separate electrical digitization and lock-in analysis channels. This approach insured those variations in the tip condition, feedback glitches or other sample or cantilever transient conditions were consistent between the measurements. All the OBD and IDS images in FIG.s 8, 10 and 11 and in FIG.s S8-S19 were acquired in this manner.

The length *L* of the cantilever was measured from the digital camera on our microscope. The position of the spot along the length of the cantilever *x* was measured relative to the base of the cantilever. The position of the base of the cantilever was determined ex post facto from the optical images. Because the base of the cantilever was out of focus due to the tilt of the cantilever, the position of the base was taken to be the point at which the contrast changed the most sharply. Position error is introduced from these measurements as well as the bending of the cantilever which is not exactly known. The combination of these uncertainties imposes absolute positioning errors of <5% for $x/L$, which is small compared to the wide discrepancies in OBD results observed in this study but sufficient to explain the relatively small discrepancies in IDS measurements (typically <2%). Note that the relative positioning errors between sequential positioning steps are much smaller, typically <1%.

### B. Method 2 -Minimum Electrostatic Force Method (MEFM)

The electrostatic force at the drive frequency is proportional to the product of the AC and DC biases. One method of searching for the electrostatic null point is to choose a test spot location, ramp the DC bias, and then record amplitude. The laser spot position for which the amplitude changes by the smallest amount is then defined as the electrostatic null point. Our approach to implementing this method was identical to the previous Equal Amplitude Method with the added step of taking multiple images at each spot position with varying DC bias values.

### C. Method 3 – Reference Sample Method (RSM)

Killgore et al. suggested another method for finding the null point where the cantilever is engaged on a charged but not piezoelectric “reference” surface. The laser position is adjusted until a minimum amplitude is observed. Then, the reference sample is removed and a piezoelectric sample inserted into the microscope and imaged using the same laser spot position. In our implementation of this method, we used fused silica as a reference material.

## IV. RESULTS

Here, we report on systematic comparative PFM measurements using the null point methods outlined above for both IDS and OBD measurements. We intentionally limited the choice of samples and cantilevers to minimize the number of uncontrolled variables as much as possible and to generate correlated statistics. We chose PPLN because it is commercially available, it has an intermediate effective piezo sensitivity with well-defined vertical domains and is well-known to have surface charge that leads to significant BES forces that will challenge the null point methods. In LNO, there is some possibility that flexoelectricity may cause a true physical amplitude asymmetries over oppositely poled domains on the order of 10%.[56] While we have assumed this flexoelectric effect is not present for the purposes of this study, we believe that the results presented here could pave the way to accurate testing of flexoelectricity in the future. We chose a single type of diamond-coated, conductive cantilever, the Adama AD2.8 because their high conductivity and superior wear characteristics and reproducibility made them ideal for systematic and long-term, repetitive studies.

### A. Method 1: Equal amplitude method - EAM

For the EAM, an automated data capture procedure where the OBD and IDS laser spot position was moved with a stepper motor along the length of the cantilever while PFM images were captured at each laser spot position was utilized as described in the Methods section. The applied force, and scan angle could be programmatically varied. Once the images were acquired, the optimized OBD and IDS null points were identified from post processing of the large data set. Our automated approach allowed for a very large dataset to be generated, in total we tested over twenty different cantilevers (of the identical make and model) and acquired >10,000 images during the course of the measurements summarized below.

FIG. 2 shows an example data set at a scan angle of $90°$, a DC tip bias of $V_{DC} = 0$, and an applied load of 50nN. After the sensitivity of the OBD detector was calibrated with a force curve, the cantilever tip was engaged on the surface at a predetermined load and images were acquired. We acquired the trace and retrace PFM amplitude images at each laser spot location. Measuring both trace and retrace curves is one common way of evaluating the consistency of AFM results. Significant disagreement between trace and retrace images is evidence of in-plane forces affecting the measurement. FIG 2(a) shows a composite image of the top of the cantilever used to make the amplitude measurements shown in (b) and (c). The OBD spot position ranged from $x_{OBD}/L = 0.15$ to $x_{OBD}/L = 0.33$, while the IDS image ranged from $x_{IDS}/L = 0.88$ to $x_{IDS}/L = 1.06$. The starting and ending points for the OBD and IDS spots were chosen to bracket the estimated null points, previously estimated with some quick measurements before setting up the systematic, automated spot position sequence. The cantilever length was estimated from a bottom photo of the cantilever tip and has an estimated positional error of ~3microns.

FIG 2(b) shows the trace and retrace amplitude images simultaneously acquired OBD and IDS images at 14 different positions along the cantilever, measured at positions shown in FIG. 2 (b). For example, the OBD data taken at $x_{OBD}/L = 0.15$ was acquired simultaneously with the IDS data taken at $x_{IDS}/L = 0.88$. The optimal locations are circled with red, solid lines for OBD and blue, dashed lines for IDS measurements in FIG. 2 (b). FIG. 2 (c) shows averaged, color-coded sections of the images in FIG. 2 (b). In this particular dataset, the EAM null point position was different for the OBD trace and retrace data; $x_{OBD}/L = 0.26$ for OBD Trace and $x_{OBD}/L = 0.27$ for OBD Retrace (red circles in FIG. 2 (b), while the IDS null point was the same for Trace, at the tip location to within our experimental error (blue circles in FIG. 2 (b), indistinguishable from the tip location ($x_{IDS}/L = 1.02$). The OBD EAM null point amplitude values, calibrated with a force curve, were significantly larger than those of the IDS, calibrated by the wavelength of light (note the different color bar ranges in FIG. 2 (b) and the corresponding y-axis ranges in FIG. 2 (c)). Measurements at the spot position extrema are designated in the plot legend.

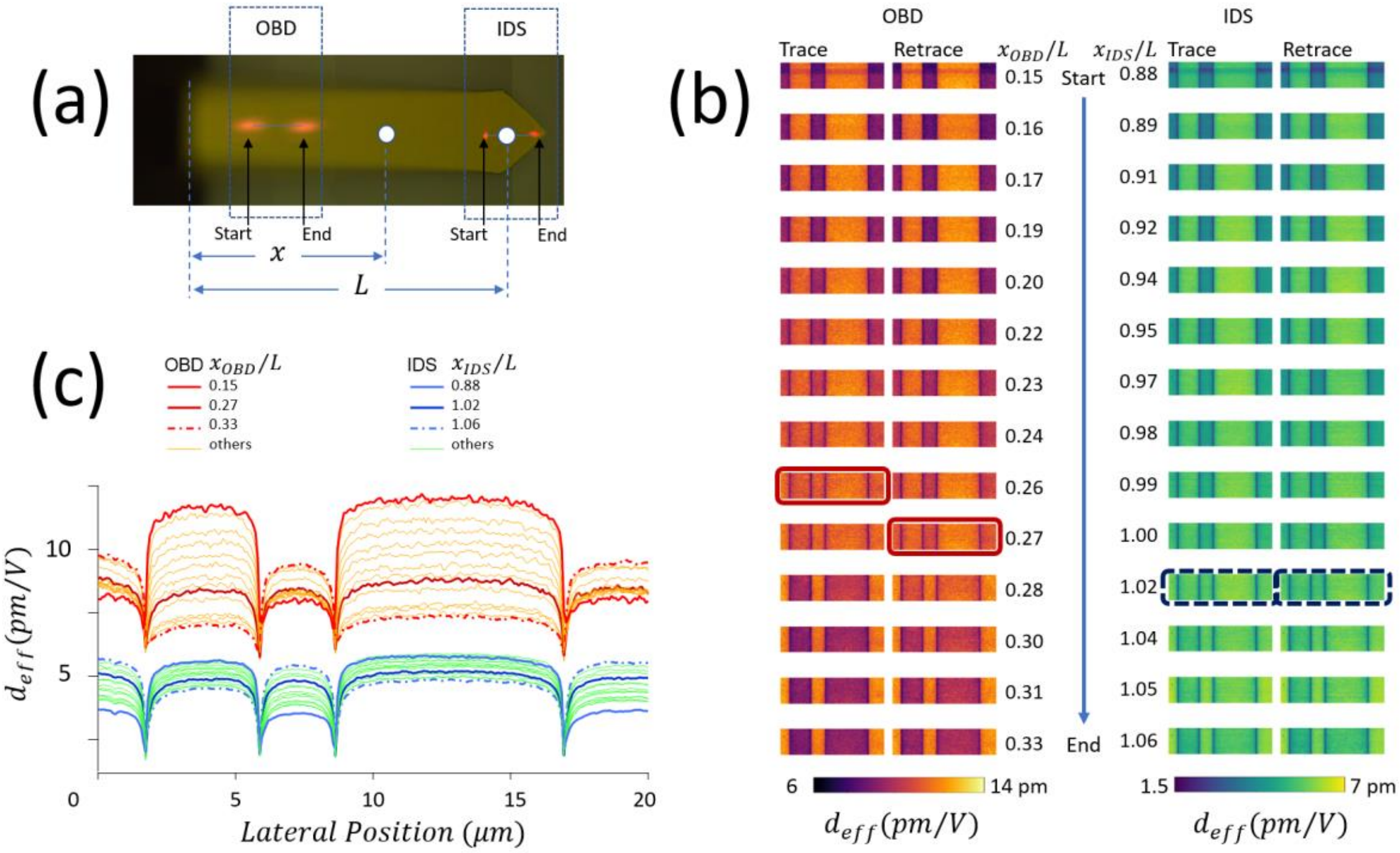

FIG. 2. EAM Results (a) The composite photo of the cantilever shows the start and end laser spot locations where OBD and IDS data was acquired. These positions were chosen to bracket the null points for each detection method. (b) shows 14 simultaneously acquired trace and retrace images with both OBD and IDS. In this panel, the optimal equal-amplitude spot locations can be picked out by inspection and are circled. (c) shows averaged retrace sections.

In our implementation of the EAM, the OBD and IDS spot positions were offset from each other, while the images using each detector were acquired simultaneously, a capability of our metrological AFM. A composite photo of a cantilever as shown in FIG. 2 (a), illustrating the locations for the OBD start and end positions and the IDS start and end positions. In this case, the starting and ending positions were chosen to bracket the “null points” for each detection method. FIG. 2 (b) shows the trace and retrace amplitude images at 14 different positions along the cantilever. Images that best fulfilled the EAM criterion were circled for each series and designated as a null point. Note that the null point position was different for the OBD trace ($x_{OBD}/L = 0.26$) and retrace ($x_{OBD}/L = 0.27$), while the IDS null point was the same for trace and retrace and was almost identically at the tip location ($x_{IDS}/L = 1.02$). (c) shows plots of image sections averaged along the vertical axis for all the images in FIG. 2 (b).

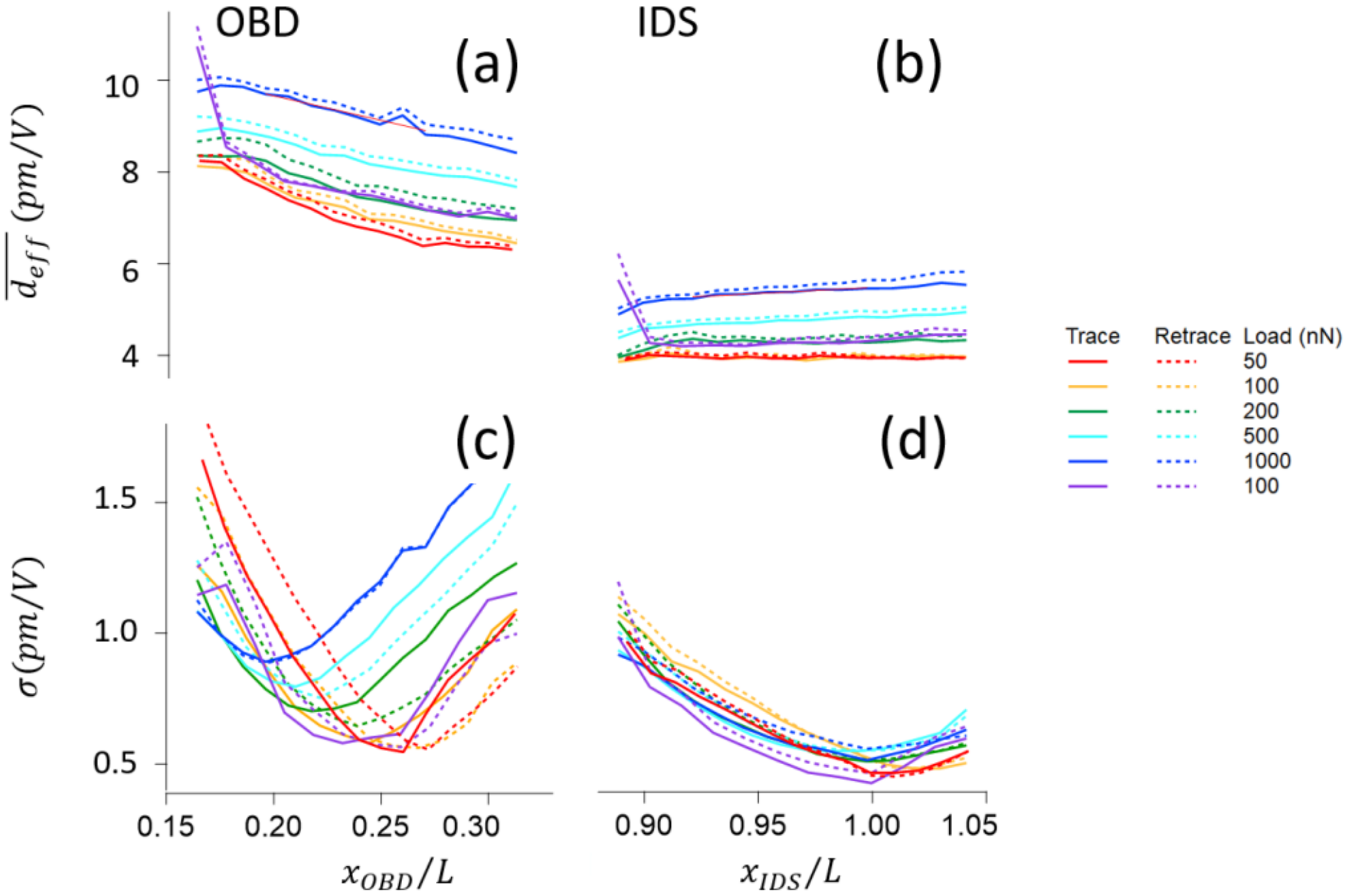

FIG. 3 (a) and (b) show the average values of the OBD and IDS amplitudes respectively versus spot position. The solid (dashed) lines are for the Trace (Retrace) images. The time sequence of loads used is in the legend, starting with 50nN (red), 100nN, 200nN, 500nN, 1,000nN and then decreased to 100nN. (c) OBD standard deviation over the entire amplitude image and (d) IDS standard deviation over the entire amplitude image. For the purposes of the discussions in this paper, the minimum value of the standard deviation is used to programmatically identify the equal-amplitude condition.

The EAM measurement procedure shown in FIG. 3 was repeated at eight different load values: increasing from 50nN, 100nN, 200nN, 500nN, 1,000nN and then decreasing to 100nN and 50nN. Averaged sections for these measurements appear in the supplemental material. Statistics from these images are shown in FIG. 3. The average and standard deviation of the amplitude from each image is plotted as a function of laser spot position for different levels of applied load. The BESNP for the EAM is the spot position with a local minimum in deviation.

From FIG. 3, we observe that both IDS and OBD measurement statistics depend on the applied load. In FIG. 3 (a), the OBD $\overline{d_{eff}}$ values ranged from 6.4pm/V to 9.8pm/V, while in FIG. 3 (b), the IDS amplitudes ranged from 3.9pm/V to 5.8pm/V. Both OBD and IDS $\overline{d_{eff}}$increased with increasing load. Normalized to the average values at the equal-domain locations, as the load changed between 50nN and 1,000nN, the OBD $\overline{d_{eff}}$ varied by 42%, while the IDS varies 39%. This result is consistent with the tip-sample stiffness increasing with the load, thereby increasing the coupling between the surface strain and the driving force on the cantilever tip (see FIG. 5 and associated discussion).

Both $\overline{d_{eff}}$ and $\sigma$ for the OBD measurements (FIG.s 9 (a) and (c)) were more strongly dependent on spot position $x$ than for the IDS (FIG. 3 (b) and (d)). For example, OBD $\overline{d_{eff}}$ in FIG. 3 (a) varied by ~20% while IDS $\overline{d_{eff}}$ in FIG. 3 (b) varied by ~5% over the scanned range of spot positions. For both the OBD and IDS measurements, the sensitivity of $\overline{d_{eff}}$ to the normalized positioning error ranged from $\partial\overline{d_{eff}}/\partial(x_{OBD}/L) \approx -17.0pm$ at the 50nN load to $\partial\overline{d_{eff}}/\partial(x_{OBD}/L) \approx -11.0pm$ for the 1,000nN load. For the IDS, $\partial\overline{d_{eff}}/\partial(x_{IDS}/L) \approx 0.06pm$ at 50nN and $\partial\overline{d_{eff}}/\partial(x_{IDS}/L) \approx 2.6pm$ at 1,000nN. For the OBD images, the average value of $\overline{d_{eff}}$ ranged from $7.2pm$ to $8.9pm$ ($\Delta\overline{d_{eff}} \approx 1.7pm$) while the IDS varied between $3.3pm$ to $3.5pm$ ($\Delta\overline{d_{eff}} \approx 0.2pm$). Referenced to the null point $\overline{d_{eff}}$, the IDS varied ~7% over the spot while the OBD varied ~23%.

FIG.s 9 (c) and (d) show the deviation ($\sigma$) of the OBD and IDS, respectively, $\overline{d_{eff}}$ measurements. We used the minima of $\sigma$ to automatically identify the optimized EAM spot position. Inspection of the minima in FIG. 3 (c) shows that the OBD EAM position was clearly dependent on load, ranging from $x_{OBD}/L \approx 0.19$ at the highest load to $x_{OBD}/L \approx 0.27$ at the lowest load. FIG. 3 (c) also shows a large variation between the trace (solid) and retrace (dashed) traces, including the minima locations. This observation was a hint that in-plane forces, in addition to long-range electrostatic forces, play a role in the location of the EAM spot location. FIG. 3 (d) shows the same deviation curves for the IDS measurements. In contrast to the OBD measurements, the minima of the IDS measurements were roughly load-independent, at $x_{IDS}/l \approx 1$, with very small differences between trace and retrace.

FIG. 4 shows a measurement we made to explore the effects of in-plane forces on the EAM spot position for both OBD (FIG. 4 (a) and (b)) and IDS (FIG. 4 (c) and (d)) measurements. First, while scanning the sample at 90 degrees (perpendicular to the longitudinal axis of the cantilever), we located the minimum $\sigma$ spot position for both OBD ($x_{OBD}/L \approx 0.33$) and for IDS ($x_{IDS}/L \approx 0.98$). Inspection of FIG. 4 (a) for the $90°$ and $270° (= -90°)$ degree scans shows roughly equal amplitudes over opposite domains while, FIG. 4(c) shows the same for the IDS amplitude, where the response is largely independent of scan direction. Note that, as discussed above, the OBD response is larger ($d_{eff}^{OBD} \approx 18pm/V$) than the simultaneously acquired IDS response ($d_{eff}^{IDS} \approx 8pm/V$).

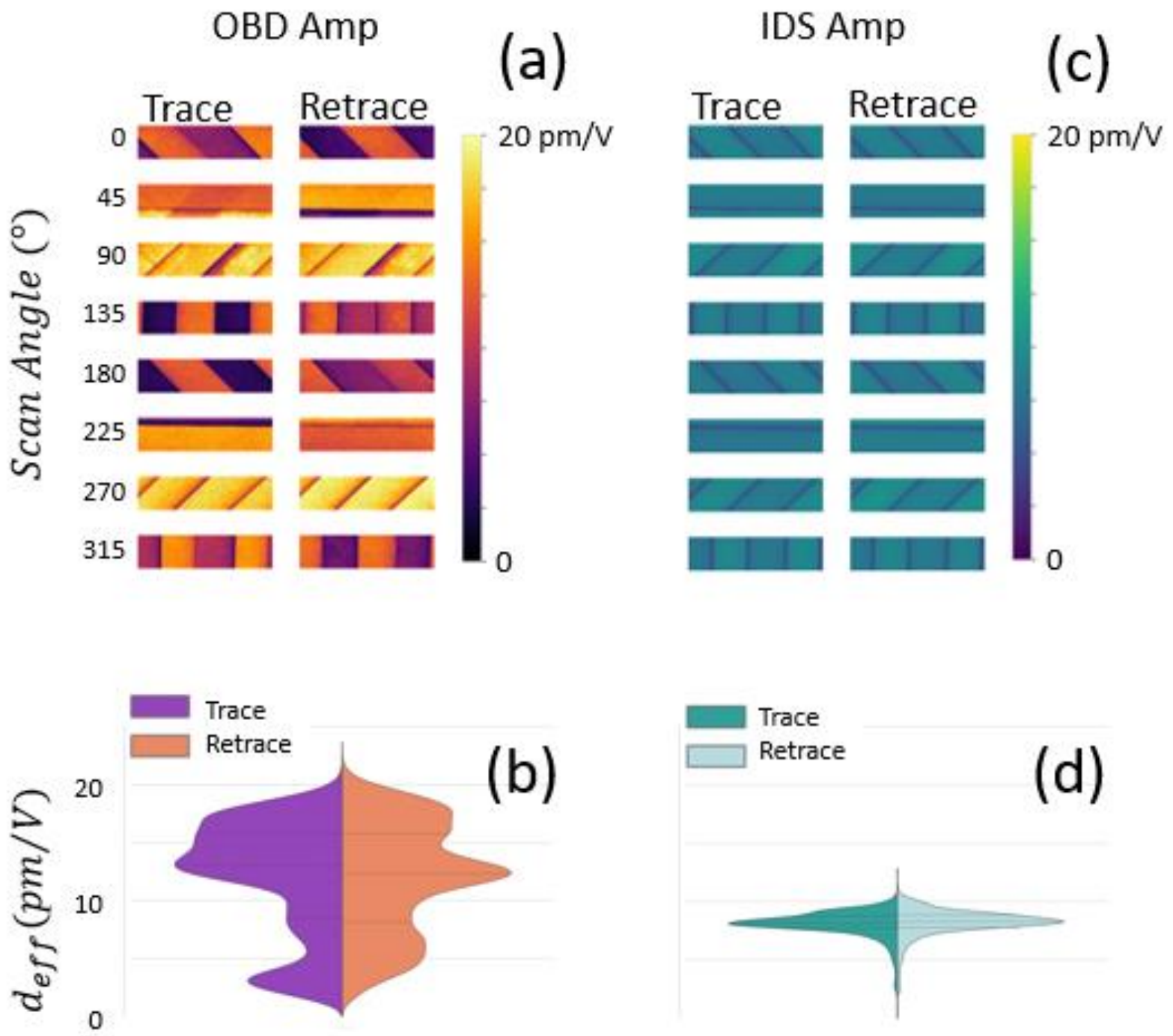

FIG. 4 (a) the trace and retrace amplitude images as a function of eight different scan angles. Note that Trace for 0 degrees should be roughly equivalent for retrace at 180 degrees. The OBD spot position, $x/L \approx 0.33$, was chosen to satisfy the EAM condition at a scan direction of 90 degrees. (b) shows the ***consolidated*** violin histogram for all the scan angles in (a). By "consolidated", we mean that the trace and retrace histograms in (b) contain the amplitude values over all the images shown in (a). (c) shows simultaneously measured IDS amplitudes at spot location $x_{IDS}/L \approx 0.98$. (d) shows the consolidated histograms for the data in (c). The data in this FIG. is the only exception to the simultaneous acquisition of data in this paper, in this case, the OBD and IDS images were acquired sequentially.

As the scan angle was changed, the IDS amplitudes in FIG. 4 (c) stayed relatively consistent whereas the OBD amplitudes in FIG. 4 (a) changed significantly. In addition, there is significant variation between trace and retrace OBD amplitude images when the scan angle differs from 90°. FIG. 4 (b) and (d) show the consolidated $d_{eff}$ histograms for OBD and IDS respectively. Both the scan angle dependence and the variation between trace and retrace images suggest that BESNP OBD is much more sensitive to in-plane forces than IDS measurements. This result also suggests that varying the scan angle is a simple and effective test for the presence of in-plane forces that may be affecting the accuracy of vertical amplitude measurements.

B. **Method 2 – Minimum electrostatic force method (MEFM)**

As discussed above and in the methods section, this method attempts to optimize the null point location by finding the spot location where variation of the measured amplitude is minimized with respect to an applied DC bias. FIG. 5 shows the effect of changing the DC tip bias for the tested sample and cantilever combination. While scanning the sample at 90 degrees and driving the cantilever with $V_{AC} = 5V$, we recorded both the IDS and OBD amplitudes at different spot positions as $V_{DC}$ was varied between $-10V$ and $+10V$ in $5V$ increments. Different laser spot positions showed different sensitivities to changes $V_{DC}$. Both OBD and IDS have clearly defined spot positions where the deviation of the average amplitude within the same PPLN domain with respect to DC tip bias was negligible. The minimum in IDS deviation happened at $x_{IDS}/L \approx 0.98$ and the minimum in OBD deviation happened at $x_{OBD}/L \approx 0.55$. These spot positions are the BESNP for the MEFM. However, when we observe the corresponding amplitude images for these spot positions, we find that OBD does not have equal amplitudes in both PPLN domains while IDS does satisfy this condition (the OBD at $x_{OBD}/L \approx 0.64$ has a 32 % difference in domain amplitudes while IDS at $x_{IDS}/L \approx 0.98$ has a 1.8 % difference in amplitudes). If we adjust the OBD spot position to $x/L \approx 0.64$, the EAM condition is satisfied at $V_{DC} = -10V$, we find that the observed response in both PPLN domains does not remain equal as the DC tip bias is increased. For OBD at $x_{OBD}/L \approx 0.64$ the variation between domains is 1.8 % at $V_{DC} = -10V$ and 62 % at $V_{DC} = 10V$. This result illustrates that for OBD the spot position where an equal domain response is observed is not always the same spot where the sensitivity to changes in the DC tip bias is minimized. The IDS data does not suffer from this same discrepancy. Because we have captured the IDS and OBD data simultaneously, we can rule out any tip or sample differences leading to these differences. We hypothesize that the unequal domain response in OBD at $x_{OBD}/L \approx 0.55$ are due to uncontrolled in-plane forces. Note that the measurements made at $x_{OBD}/L \approx 0.64$ and $V_{DC} = -10V$, were the only dataset where we were able to substantially duplicate the results of Killgore et al., meaning that both the EA condition was met and where $d_{eff}^{IDS} \approx d_{eff}^{OBD}$. We speculate that this is because we had canceled the background electric field and that at $x_{OBD}/L \approx 0.64$, we were near the longitudinal null point for the cantilever.

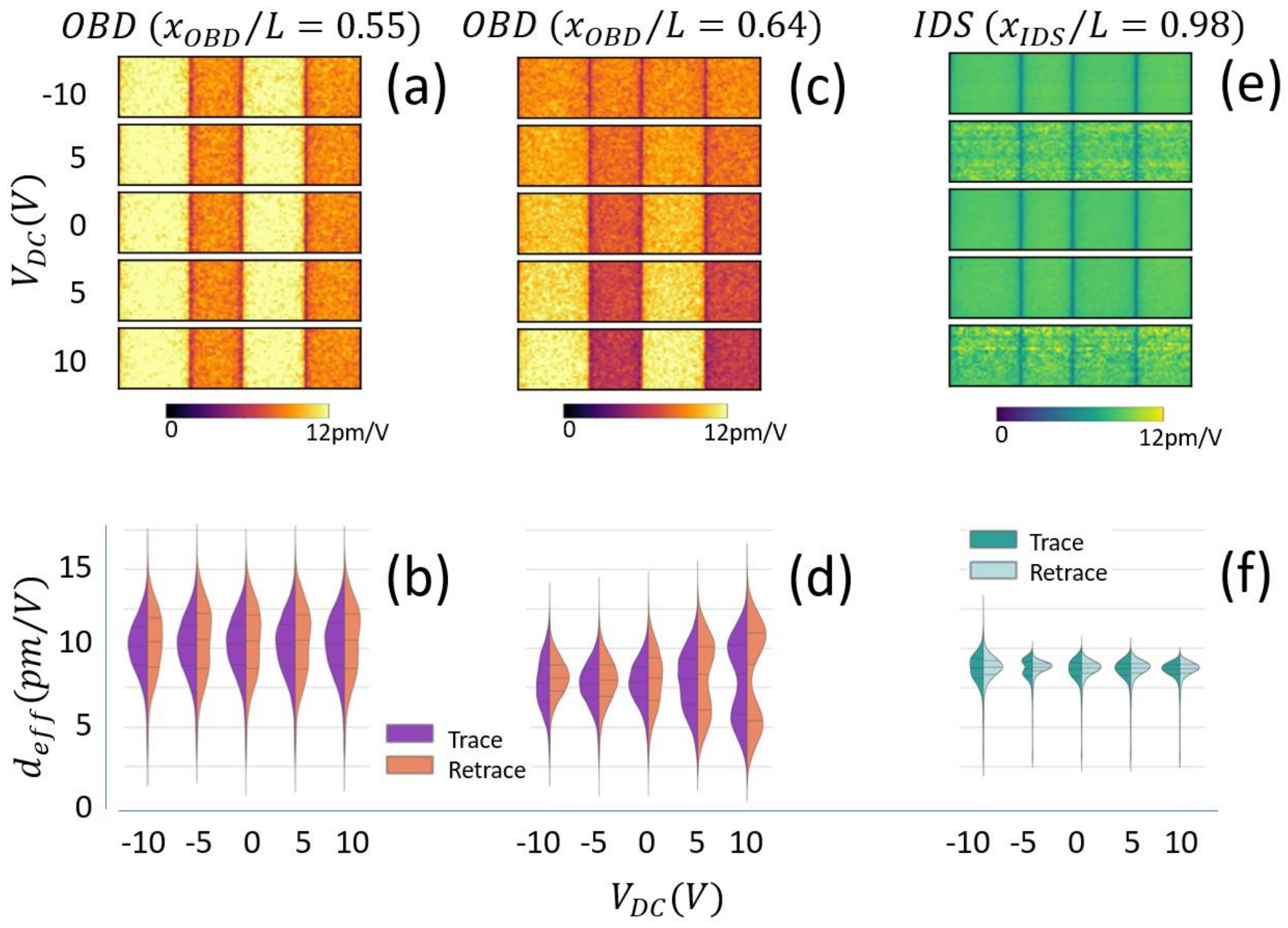

FIG. 5. Variation in response as a function of DC tip bias. (a) OBD PFM amplitude images at a spot position of $x_{OBD}/\mathrm{L} \approx 0.55$. This is the OBD spot where the variation in single domain amplitude with respect to DC tip bias is minimized. (b) shows the associated violin histograms. (c) OBD PFM amplitude images at a spot position of $x_{OBD}/\mathrm{L} \approx 0.64$. This is the OBD spot where PPLN has equal response in both domains at $\mathrm{V_{DC}} = -10\mathrm{V}$ . (d) shows the associated violin histograms. (e) IDS PFM amplitude images at a spot position of $x_{IDS}/\mathrm{L} \approx 0.98$. This IDS spot has equal amplitudes in both PPLN domains and a minimum in sensitivity to DC tip bias. (f) shows the associated violin histograms.

### C. Method 3 – Reference sample method (RSM)

In this method, the null point position is chosen by minimizing the electrostatically driven amplitude on a non-piezoelectric sample and then, without changing the spot position, using the same cantilever to measure a sample of interest. We used clean fused silica as our reference material. Data from six separate RSM trials with different cantilevers are shown in FIG. 6, all acquired with a load of 500nN. FIG. 6 (a) shows the OBD trace and retrace amplitudes, (c) shows the IDS trace and retrace amplitudes. (b) shows the spot locations for the OBD (Blue color, larger spot) and IDS (red, smaller). FIG. 6 (d) and (e) show violin distribution plots of both the trace and retrace data for the OBD and IDS, respectively. Five out of six of the OBD data sets violated the equal-amplitude condition, with clearly separated amplitude distributions over different domains for both trace and retrace. This separation of different domains was only present in one of the IDS data sets (Trial #3).

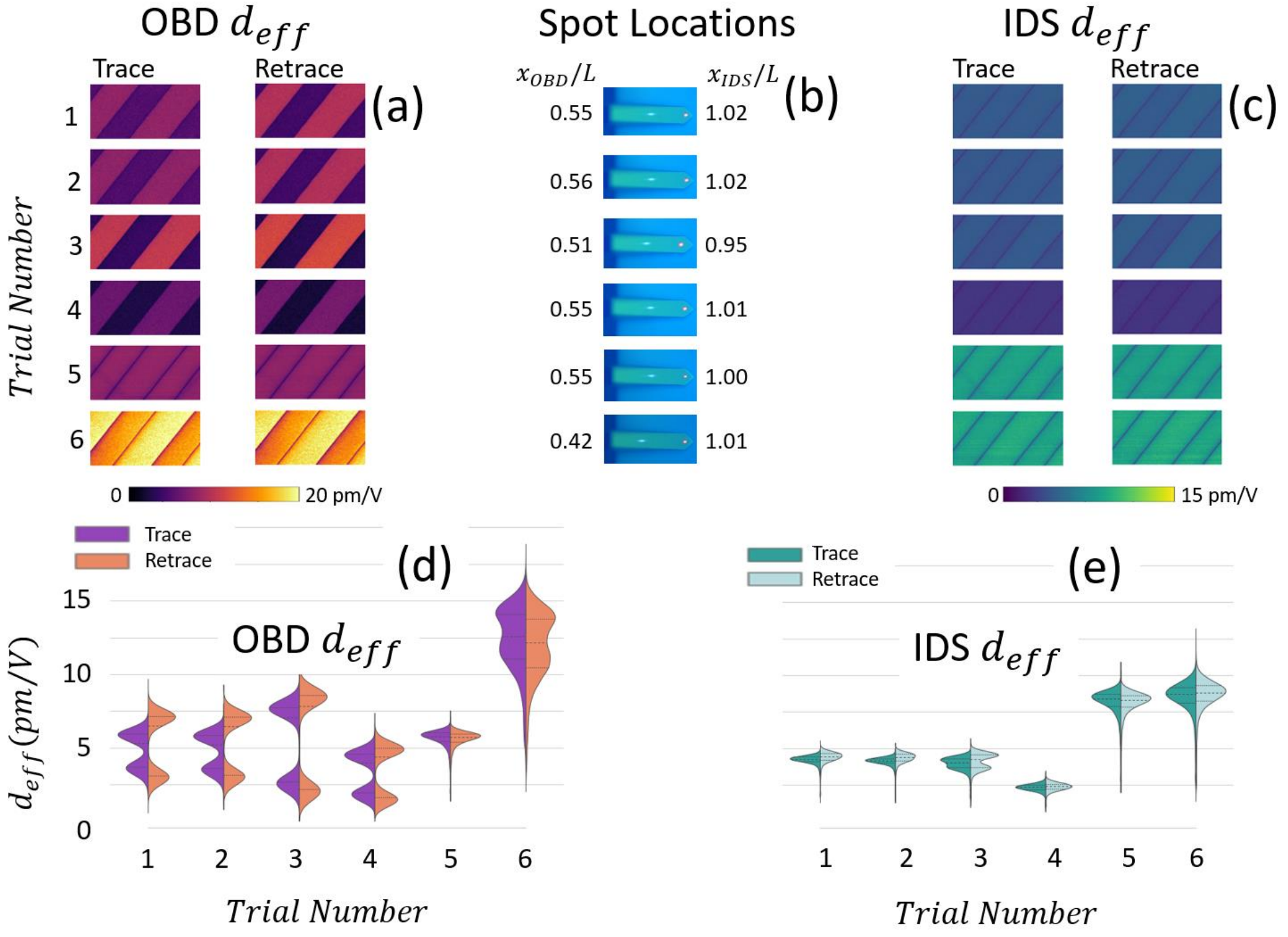

FIG. 6 (a) OBD trace and retrace amplitude images, (b) optical images showing the OBD and IDS spot locations that minimized the electrostatic amplitudes over a fused silica surface, (c) IDS trace and retrace amplitude images (d) violin distribution plots for the OBD images in (a) and (e) violin distribution plots for the IDS images shown in (c).

**II. THEORETICAL METHODS**

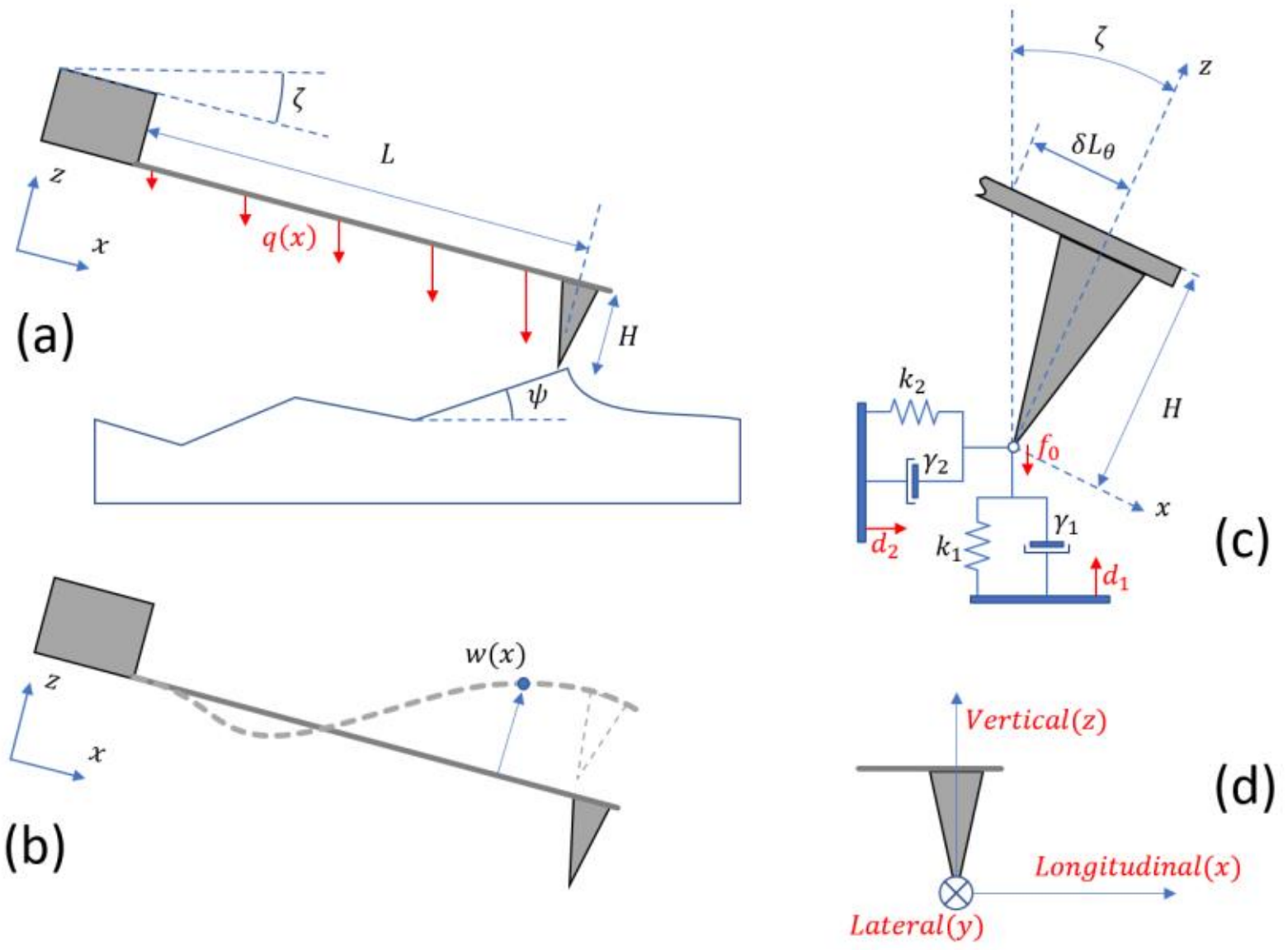

FIG. 7 (a) shows the E-B model geometry of the model, adapted from Bradler. (b) shows the displacement of the Euler-Bernoulli lever displacement $w(x)$. (c) shows a more detailed model of the tip-sample interaction parameters and some geometrical factors. (d) illustrates the definitions for vertical, longitudinal and lateral directions in the text.

We studied the cantilever response by implementing the Euler-Bernoulli solution originally presented by Bradler et al. [57], [58], [59] FIG. 7 is an adaptation from Bradler.[58] The model includes the effects of normal and longitudinal tip-sample stiffness $k_1$ and $k_2$ and damping $\gamma_1$ and $\gamma_2$, the surface normal and longitudinal sample displacements $d_1$ and $d_2$. FIG. 7(a) shows the orientation of the cantilever to the sample along with the electrostatic force acting on the cantilever body $q(x)$ and a localized electrostatic force acting at the cantilever $f_0$. The calculations were all performed in a Python 3.6 environment.

Table I shows the calculation parameters for FIG.s 2 and 3. In each of the panels in FIG. 2, the total beam motion is the sum of the vertical motion (black), the in-plane forces (red) and the electrostatic forces (orange). The dashed lines refer to negative vertical motion while the solid lines are for positive vertical motion. The green dots on the panels represent spot positions (values of x/L) where the total response gives the correct IDS displacement ((a) and (c)) values or the correct OBD slope ((b) and (d)) values.

The total motion of the Euler-Bernoulli beam is the linear sum of the independent contributions. As is shown in the orange and red curves in FIG. 8, other components of the polarization vector can contribute to the vertical PFM signal through a combination of vertical and longitudinal bending. FIG. 8 shows two different examples of the addition of these contributions, (a) and (c) show the displacement, w(x) and measured with an IDS, while (b) and (d) show the associated slopes w'(x), measured with the OBD.

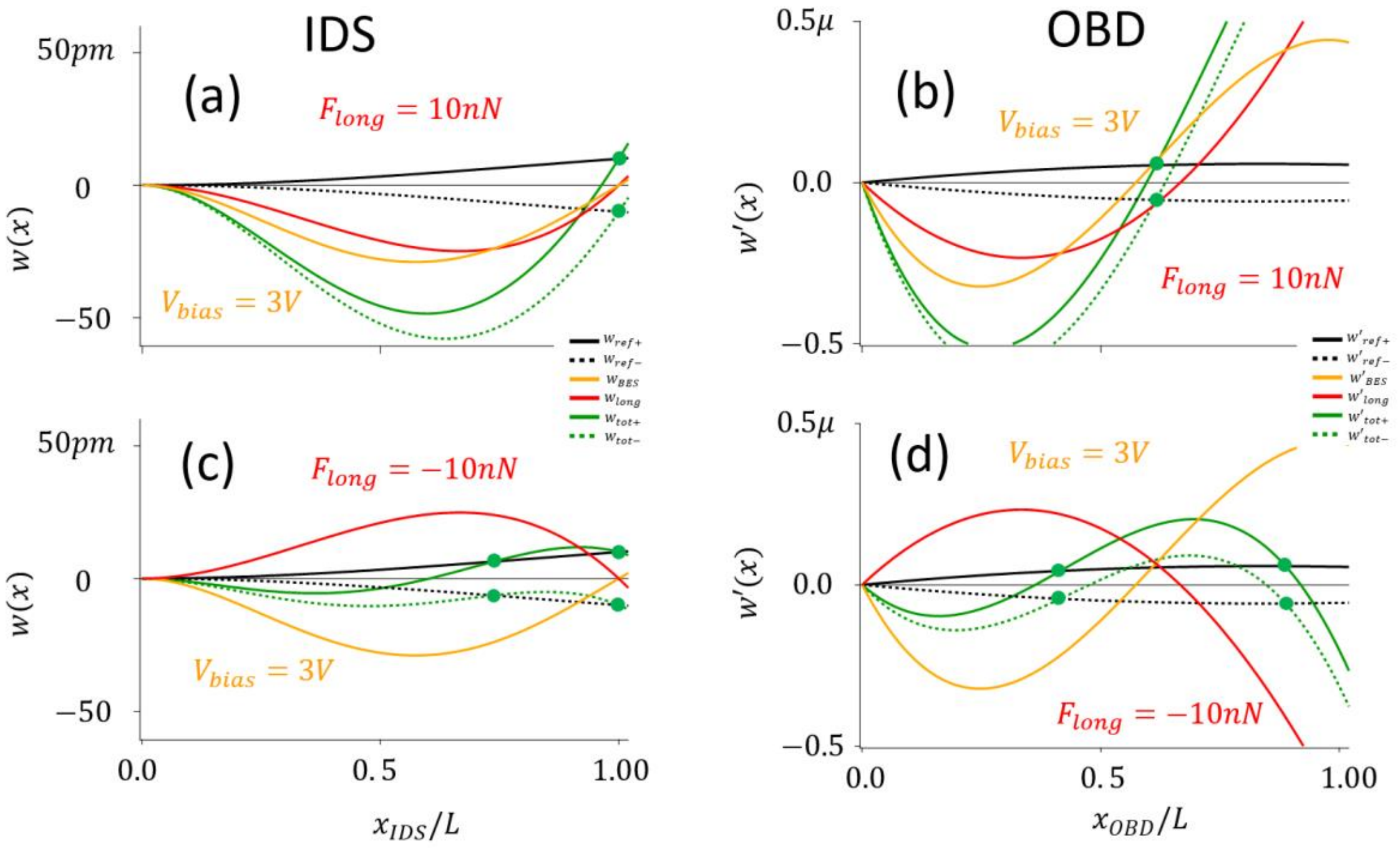

FIG. 8 (a) and (c) Euler-Bernoulli cantilever beam displacement ($w(x)$) curves as measured with the IDS versus $x_{IDS}/L$ and (b) and (d) deflection (slope, $w'(x)$) as measured with OBD versus $x_{OBD}/L$. In all four panels. the black solid lines show the cantilever response to a purely ***positive*** vertical excitation, where the tip is allowed to freely slip along the surface (kx=0); the dashed black lines show the purely ***negative*** vertical excitation. In all four panels, the orange curve shows the response to only BES forces while the red curves show the response to (a) and (b), a ***positive longitudinal*** force and (c) and (d) a ***negative longitudinal*** force. The green solid (dashed) curves show the total linear sum of the positive (negative) vertical, longitudinal and BES motions. Points where the total motion intersect the black curves (denoted with green circles) are the spot positions where BES and longitudinal contributions sum to zero, leaving only the vertical response. These are the spot positions where the detector sensitivity to the BES and longitudinal responses vanishes and only the vertical response sensitivity remains. More details are listed in Table I.

Table I: Simulation parameters for FIG.s 8 and 9.

| $k_1(N/m)$ | $k_2(N/m)$ | $d_1(\mathrm{m})$ | $d_2(\mathrm{m})$ | $V_{DC}(\mathrm{volts})$ | FIG. 8 | FIG. 9 | Legend |
|---|---|---|---|---|---|---|---|
| $10^5$ | 0 | $10^{-11}$ | 0 | 0 | $w_{ref+}, w'_{ref+}$ | $S_{ref+}, S'_{ref+}$ | solid black |
| $10^5$ | 0 | $-10^{-11}$ | 0 | 0 | $w_{ref-}, w'_{ref-}$ | $w_{ref-}, w'_{ref-}$ | dashed black |
| $10^5$ | $10^2$ | 0 | $-10^{-12}$ | 0 | $w_{ES}, w'_{ES}$ | NA | solid red |
| $10^5$ | $10^2$ | 0 | 0 | 10 | $w_{long}, w'_{long}$ | NA | solid orange |
| $10^5$ | $10^2$ | $10^{-11}$ | $-10^{-10}$ | 10 | $w_{tot+}, w'_{tot+}$ | $S_{tot+}, S'_{tot+}$ | solid green |
| $10^5$ | $10^2$ | $-10^{-11}$ | $-10^{-10}$ | 10 | $w_{tot-}, w'_{tot-}$ | $S_{tot-}, S'_{tot-}$ | dashed green |

Because the response of the Euler-Bernoulli beam can vary enormously in response to different boundary conditions, it is convenient to introduce a “shape factor” that allows easy and consistent comparison of the cantilever shape in response to various unwanted or unintended stimuli with the ideal response in the presence of only the interaction of interest. In this case, that is a vertical strain in the absence of any local or global electrostatic forces and in the absence of any longitudinal forces. If we denote the cantilever displacement subjected only to the target interaction as $w_{ref}(x)$ and the cantilever displacement in the presence of all the forces as $w_{tot}(x)$ we can then define the IDS displacement shape factor of the lever as

$$S_{IDS}(x) = \frac{w_{tot}(x)}{w_{ref}(x)}. \quad (1)$$

Similarly, we can define the OBD slope shape factor as

$$S_{OBD}(x) = \frac{w'_{tot}(x)}{w'_{ref}(x)}. \quad (2)$$

This approach is convenient in many ways since it allows direct comparison of displacement and slope-based detection approaches on the same plot. We can trivially define the shape factor boundary conditions as

$$S_{ref,\ IDS}(x_{IDS}) = \frac{w_{ref}(x_{IDS})}{w_{ref}(x_{IDS})} = 1 \quad (3)$$

and

$$S_{ref,\ OBD}(x_{OBD}) = \frac{w'_{ref}(x_{OBD})}{w'_{ref}(x_{OBD})} = 1. \quad (4)$$

Note that while we lose sign information with this convention, meaning that both positive and negative vertical domains map to unity, it allows very easy quantitative comparisons between different scenarios, including IDS and OBD detection modes. Because of this, we will simplify the notation to $S_{ref,\ IDS}(x_{IDS})$=$S_{ref,\ OBD}(x_{OBD}) = S_{ref} = 1$. The points where $S_{IDS}(x) = 1$ or $S_{OBD}(x) = 1$ are null points, where the measured motion is identical to $w_{ref}(x)$ for the IDS or identical to $w_{ref}'(x)$ for the OBD. If the detector spot is located at those positions, only the vertical response of the sample will be measured.

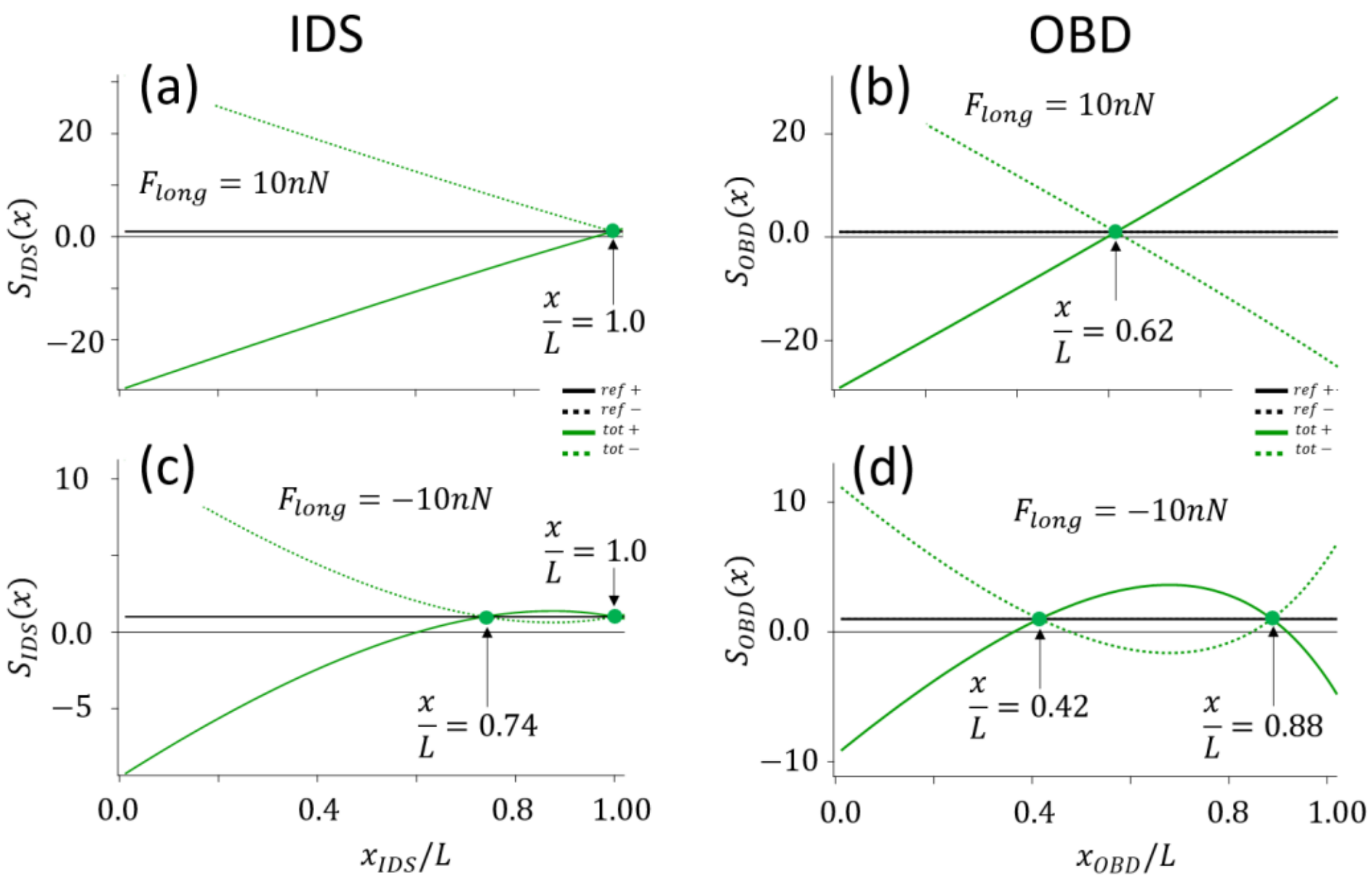

FIG. 9. Shape factors for the examples in FIG. 8. The shape factors (green) were calculated with (1) and (2). There is a horizontal black line at $S_{ref} = 1$ in all the FIG.s. The points where the shape factors (green solid and dashed lines) intersect $S_{ref} = 1$ are the null points for the excitation parameters. The parameters are listed in Table I and in the legend. As with FIG. 8, the IDS shows a null point at $x_{IDS}/L = 1$ in both (a) and (c), showing that the null point is substantially immune to changes in the longitudinal forces in contrast to the OBD null points.

Accurate theoretical interpretation of a measurement depends on calibration with the correct reference shape. The above examples shown in FIG.s 8 and 9 were calibrated with a reference shape that assumed the cantilever tip with free to slide along the sample surface ($k_x = 0$). In most experiments, we do not expect this limit to hold; friction and stiction are expected to play a role. To investigate this, we introduced the possibility of longitudinal forces, both for the reference calibration step and for the measurement step.

There are four permutations of the two steps that we label as "reference condition-measurement condition" where the first label refers to the force curve step and the second to the imaging step as follows: (i) stick-stick, (ii) slip-stick, (iii) stick slip and (iv) slip-slip.

The reference shapes (illustrated in FIG. 10) are calculated for the slipping case as $w_{ref\pm,slip} = w(x_{IDS}/L\,, \pm\,\delta z\,, V_{dc} = 0, k_x = 0, \delta x = 0)$ for the IDS and $w'_{ref\pm,slip} = w'(x_{OBD}/L\,, \pm\,\delta z\,, V_{dc} = 0, k_x = 0, \delta x = 0)$ for the OBD case. For the sticking case, the reference shapes are calculated as $w_{ref\pm,stick} = w(x_{IDS}/L\,, \pm\,\delta z\,, V_{dc} = 0, k_x = 10^4\,N/m\,, \delta x = 0)$ for the IDS and $w'_{ref\pm,stick} = w'(x_{OBD}/L\,, \pm\,\delta z\,, V_{dc} = 0, k_x = 10^4\,N/m\,, \delta x = 0)$ for the OBD. These shapes were used to define the shape factors as discussed above in the context of force curves that are either slippery, ($k_x = 0$) or sticky ($k_x = 10^4 N/m$). In addition, since force curves are mechanically driven, the reference shapes were calculated assuming $V_{DC} = 0$. Finally, to mimic a slow force curve, the references were also made at a low sub-resonance frequency, typically $\omega = 2\pi \cdot 1kHz$.

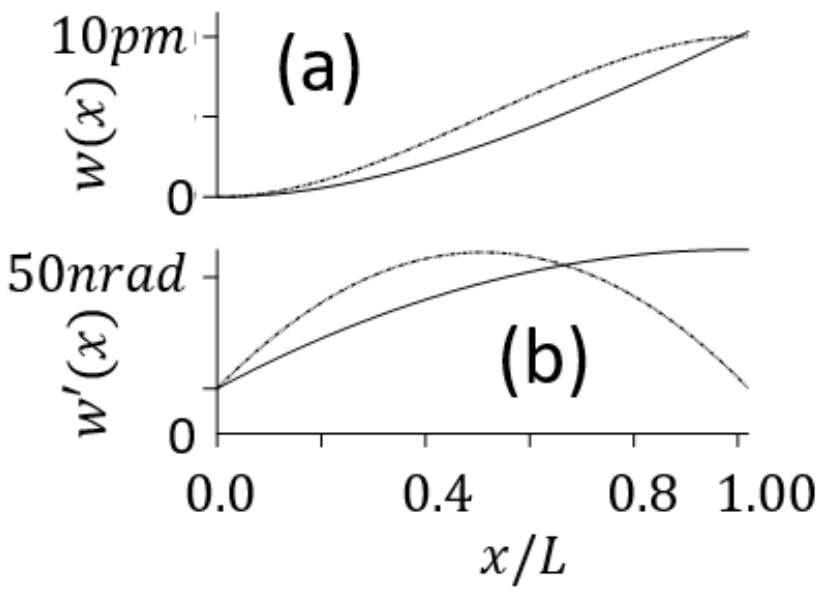

c

FIG. 10 shows the (a) displacement ($\mathrm{w(x)}$) for a slipping cantilever (solid line, $\mathrm{k_x} = 0$) and for a sticky cantilever (dashed line, $\mathrm{k_x} = 10^4 \mathrm{N/m}$) subject to 10 picometer vertical and no (0) in-plane (longitudinal) strains. (b) shows the deflection ($\mathrm{w'(x)}$) as measured with an OBD detector under the same conditions. More details are given in Table II.

In many cases, experimentalists have assumed that the equal-amplitude condition is a sufficient condition to ensure accuracy of the measurement. To simulate this condition, calculate two different shape factors, one for an up domain with strain $\delta z$ and another for a down domain with strain $-\delta z$. We then estimate null point positions by calculating the spot positions where these shape factors are equal: $S(x/L, \delta z, V_{dc}, k_x, \delta x_+) = S(x/L, -\delta z, V_{dc}, k_x, \delta x_-)$. In this expression, $\delta x_+$ and $\delta x_-$ are longitudinal strains that can be varied to explore different longitudinal symmetry assumptions as described in Table II and FIG.s 15 and 16.

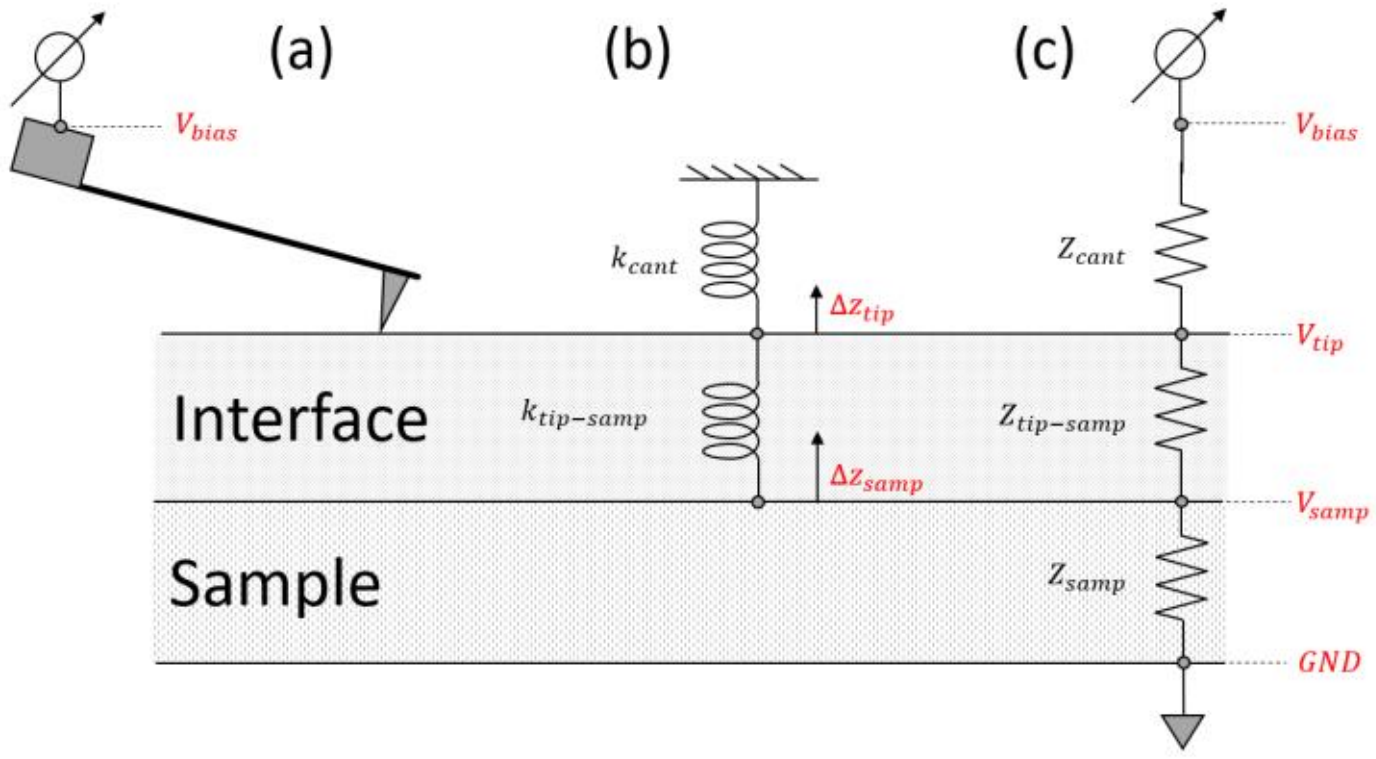

FIG. 11 (a) shows a biased probe in contact with a non-ideal sample with an interface. (b) this interface will lead to a tip-sample stiffness that reduces the motion of the tip. (c) In addition, this interface can have a non-zero impedance that may lead to a reduced bias at the true sample surface. The cantilever tip makes contact with the top of the sample surface. In ambient conditions, there will be an interface, such as a water layer, contaminants, or an oxide layer. As depicted in FIG. 11(b), the stiffness of the tip interface contact, denoted by $k_{tip-samp}$, is dependent on the contact area with this interface. In this model, we assume that the sample stiffness is significantly greater than the two stiffnesses shown in FIG. 11(b), and therefore, we neglect it. This assumption is reasonable for the PPLN sample studied in this work. If $k_{tip-samp} \gg k_{cant}$, the cantilever stiffness, the deflection of the cantilever, $\Delta z_{cant}$ is similar to the actual sample surface motion, $\Delta z_{samp}$. If the tip-sample stiffness no longer fulfills this criterion and the cantilever deflection is reduced by a factor

$$d_{eff}^{meas} = d_{eff}^{samp} \frac{k_{tip-samp}}{k_{tip-samp}+k_{cant}}. \quad (5)$$

In this Equation, when $k_{tip-samp} \to 0$, the sample motion is completely absorbed in the soft tip-sample stiffness and the measured cantilever motion also tends to zero. This simple model has some significant consequences; although the high spatial resolution of PFM is dependent on a relatively small contact area, a small contact area may also result in a reduced contact stiffness which then underestimates the sample response.

In a similar manner, FIG. 11 (c) shows the effect of finite resistance at the different components of the cantilever interface sample junctions. The impedance of the cantilever chip and tip assembly is designated $Z_{cant}$, the surface interface impedance is designated as $Z_{tip-samp}$ and the impedance of the sample itself is designated as $Z_{samp}$. As in the case of the springs in FIG. 11 (b), these impedances act as a voltage divider that reduces the bias magnitude at the surface of the sample by the factor

$$V_{samp} = V_{bias} \frac{Z_{cant}+Z_{tip-samp}}{Z_{cant}+Z_{tip-samp}+Z_{samp}} . \quad (6)$$

In the limit that that $Z_{samp} \to \infty$, the potential the sample surface experiences is the bias applied to the cantilever chip by the PFM control electronics, $V_{samp} = V_{bias}$. However, if $Z_{samp}$ is finite, $V_{samp} < V_{bias}$, in turn reducing the electromechanical response.

The above analysis is a greatly simplified picture of the tip-sample interactions that neglect the cantilever dynamics, long-range electrostatics and vector electromechanics. Nevertheless, it implies that contact area variations, since they directly affect both $k_{tip-samp}$ and $Z_{tip-samp}$, will also lead to variations in the measured electromechanical response. More specifically to repeatable, accurate measurements, it implies that high load measurements, by increasing the tip-sample contact area will favor large $k_{tip-samp}$ and small $Z_{tip-samp}$, and should thus lead to more accurate characterization of the local electromechanical response. FIG. 12 below shows both a plot of Equation (5) (red circles) and the E-B solution at x/L=1 for a deflected cantilever beam. Both calculations were made with $k_{cant} = 2.5N/m$. Note that to get within a few percent of the expected response requires $k_{ts} \gg k_{cant}$.

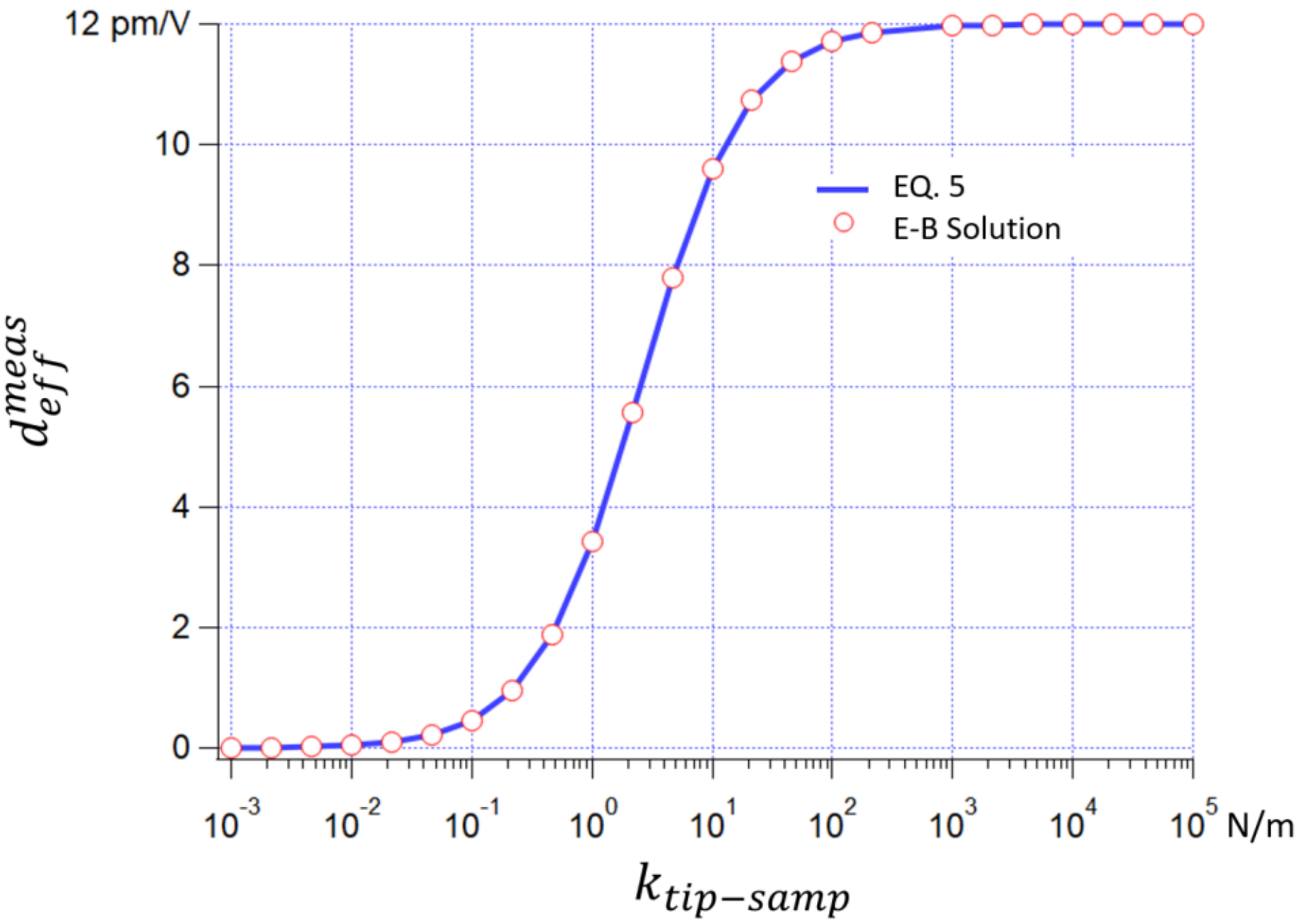

FIG. 12. Shows excellent agreement between EQ. (5) and the Euler-Bernoulli displacement solution for the IDS spot immediately above the tip, $w(x_{IDS}/L = 1)$.

The Euler-Bernoulli (E-B) model (see materials and methods) can be used to explore the location of null points for both electrostatic and longitudinal excitations. FIG. 13 shows shape factors for both OBD and IDS detection measurements. Shape factors are defined in (1) and (2) above. For the OBD measurements in this work, this reference shape is characterized with a force curve. FIG. 13 (a) shows the shape factors for a constant vertical motion ("normal", 10pm) and a range of DC bias values. The BESNP for OBD detection is $x_{ESNP,OBD}/L \approx 0.58$, while for the IDS it is $x_{ESNP,IDS}/L \approx 1.0$. Note that the null point is independent of the magnitude or sign of the DC bias value. This is consistent with the electrostatically driven mode being linear. Similarly, (b) shows the shape factors as a function of a constant, 10pm vertical ("normal") excitation and a range of longitudinal forces. In this case, the longitudinal null point, LNP for OBD detection is $x_{LNP,OBD}/L \approx 0.67$ and the LNP for the interferometer is $x_{LNP,IDS}/L \approx 1.0$. Again, as in the case of the BESNP detection, the null points $x_{LNP,OBD}$ and $x_{LNP,IDS}$ are independent of the magnitude of the longitudinal force. There is a key difference between the null points for the OBD and IDS. Whereas the IDS null point locations are identical, is $x_{LNP,IDS}/L = x_{ESNP,IDS}/L$, that is not the case for the OBD. The locations of the OBD BESNP and LNP differ by almost 10% of the cantilever length. For the nominally 225 micron levers we used here, that means that $x_{LNP,OBD}$ and $x_{EBNP,OBD}$ are ~20um apart. It also implies that when the electrostatic shape and the longitudinal shape are combined, the location of a zero-amplitude location will depend on the relative signs and magnitudes of those two components. On the other hand, in the case of IDS, since the zeros are in the same location, immediately above the tip at $x_{ESNP,IDS} = x_{LNP,IDS} = L$, even if the signs and magnitudes of the electrostatic and longitudinal components are very different elsewhere, they still sum to zero at the tip location.

In addition to the normal vertical strain, there are other excitations that lead to different cantilever shapes, specifically electrostatic and longitudinal. The cantilever shapes for vertical ($\delta z$), longitudinal ($\delta x$) and electrostatic ($V_{ac}, V_{dc}$) excitations are different. At a particular pixel in a PFM image, the cantilever shape will be determined by the linear sum of those three mode shapes. As the cantilever scans along the sample and the boundary conditions change, so does the resulting cantilever shape.

Depending on the position on the cantilever, vertical oscillations of the cantilever occur in response to longitudinal surface motion and to long range body electrostatic forces (BES). These excitations are problematic in that they directly couple into bending of the cantilever that mixes with the vertical PFM signal. FIG. 13 shows that the shape factors for OBD ($S_{OBD}$) and for IDS ($S_{IDS}$) measurements in the presence of an oscillating vertical strain force $F_z = k_{tip-samp}A_z\cos(\omega t)$, where $k_{tip-samp} = 10^6 N/m$, $A_z = 10pm$ and $\omega = 2\pi \cdot 1kHz$, applied at the tip location, $x_{IDS}/L = 1$. These parameters were meant to simulate a sub-resonant, very stiff contact electromechanical excitation similar to the expected response from LNO. (a) shows a reference line (black) that represents the response to only the vertical excitation $F_z$ (see FIG. 7 for definitions of forces acting on the cantilever and tip). The colored curves represent the response of the beam to $F_z$ plus various electrostatic excitations. Since the EB OBD solution is linear, the zero location remains constant, independent of the drive magnitude and is located at $x_{OBD}/L \approx 0.58$. This is similar to values reported by Nath et al. and Killgore et al. The same analysis for the IDS shape factor ($S$) shows the zero location for the electrostatic mode at $x_{IDS}/L \approx 1.0$, consistent with Labuda and Proksch.[40] FIG. 13 (b) shows the same analysis in the presence of in-plane (longitudinal) forces and the same vertical force as (a). The magnitude and sign of the longitudinal force are given by $R = F_x/F_z$. The $F_x$ OBD null point, $S_{OBD} \approx 0.67$ in FIG. 13 (b). As mentioned earlier, both Alikin et al.[17] and Nath et al.[16] reported that the longitudinal null point was at $x_{IDS}/L_{OBD,\ Alikin} \approx 0.59$ and $x_{OBD}/L_{OBD,\ Nath} = 141/225 \approx 0.63$ respectively. These values are significantly different from the $x_{OBD}/L \approx 0.67$ in (b). Closer reading of Alikin et al. show that they chose a different coordinate system where the tip is positioned at $x/L = 0.9$ rather than unity as we have here. We can transform Alikin et al.'s estimate of the longitudinal null point into ours by $x_{OBD}/L_{OBD,\ Alikin} = 0.59/0.9 \approx 0.66$, in close agreement to the null point in FIG. 13(b). In the case of Nath et al., there is less information available on the cantilever, but assuming it is an etched Si probe, the tip typically is offset underneath the probe. If we invoke a tip offset in addition to uncertainty in the overall cantilever length ($225\mu m$ is a round number, often reported by manufacturers as a nominal value), Nath et al. may be consistent with the OBD null point in (b). For the IDS simulations (dashed lines in (b)), the $F_x$ null point is at $x/L = 1$, at the same location it is for the electrostatic interactions in (a).

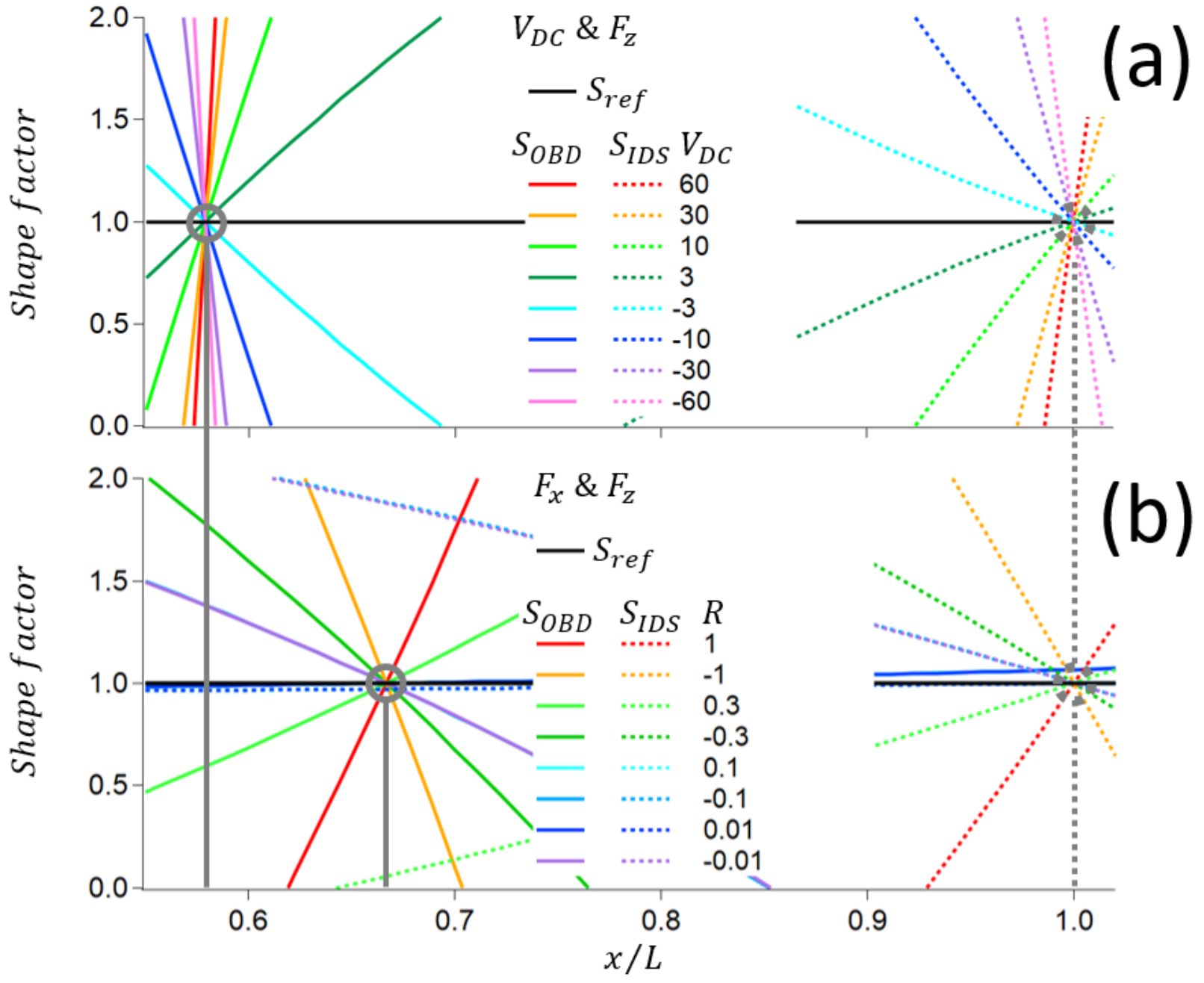

FIG. 13. (a) shows the shape factors S for the IDS and S′ for the OBD as a function of the normalized spot position x/L in the presence of a range of electrostatic biases. In all cases, the BESNP is at $x_{IDS}/\mathrm{L} \approx 1$ for the IDS and at $x_{OBD}/\mathrm{L} \approx 0.58$ for the OBD detectors. (b) shows the same analysis for a range of positive and negative longitudinal forces. In this case, the LNP for the IDS is the same as the BESNP, at $x_{IDS}/\mathrm{L} \approx 1$, while for the OBD it is at $x_{OBD}/\mathrm{L} \approx 0.67$.

In the next section, we will discuss the implications of these results. For purposes of that discussion, we break the measurement process into two steps. Step #1: force curve calibration and step #2: imaging. In step #1 the cantilever is mechanically pressed a known distance into a stiff surface while the deflection or displacement signal is measured. The slope of the deflection or displacement versus the distance the base of the cantilever moved gives of value for the detector sensitivity. The longitudinal (in-plane) forces during this step are generally poorly quantified. It is well-known that they can affect the observed signal and to demonstrate that here, we denote two different conditions – slip ($F_x = 0$) or stick ($k_x = 10^4 N/m$).[60],[61] FIG. 3 shows the shape of a cantilever that is pressed into a normally orientated surface in these two cases. FIG. 3 (a) shows the displacement as measured with an interferometer as a function of the longitudinal spot position while FIG. 3(b) shows the angular deflection as measured with an OBD. Note that in the cantilever deflection displacement measurement total distance traveled in either case with the displacement sensor at the location ($w(x_{IDS}/L = 1, \delta z = 10pm\,, k_x = 0) = w(x_{IDS}/L = 1, \delta z = 10pm\,, k_x = 10^4) = 10pm$), the distance input into the model. As you can see in both cases there are significant differences between the stick (dashed) and slip (solid) IDS displacement and OBD deflection shapes. We expect that real experimental conditions, where the cantilever tip experiences a range of in-plane forces that depend on the sample roughness, cleanliness, load and humidity[62] will be considerably more nuanced and variable than these simple limits. In addition, we limited the simulations in this work to the 0-degree tilt (cantilever parallel to the surface) limit.

FIG.s S1-S7 shows the effects of this for both OBD (left) and IDS (right) measurements. The colored traces represent a variety of shape factors for an OBD, Euler-Bernoulli cantilever in the presence both BES and in-plane forces in addition the vertical forces. The sign of the vertical force is designated with $\delta z > 0$ ($\delta z < 0$)

for positive, solid (negative, dashed) color-coded curves in all of the FIG.s S1-S7. The solid markers designate spot positions to guide the eye where the solid and dashed lines cross; these are the equal-amplitude (EA) locations. These intersections were used to calculate the EA points and are summarized in FIG. 15. When interpreting FIG. 15, refer to Table II and the symmetry conditions illustrated in FIG. 14, described below.

The different models we considered in explaining the experimental measurements are listed in Table II. The pink circle and purple square denote the "slippery" imaging conditions ($k_x = 0, \delta x = 0$, see FIG. 14(c)). The dashed blue line denotes an in-plane, longitudinal response proportional to the vertical response ($k_x = 10^4, \delta x = \alpha\delta z$, see FIG. 14 (a)). The dash-solid green and red dashed curves denote a longitudinal response proportional to the absolute value of the vertical response ($k_x = 10^4, \delta x = \beta|\delta z|$, see FIG. 14 (b)) The EA shape factors for these models are plotted in FIG. 13. Examples of the families of E-B solutions used to estimate S* for the OBD measurements are shown in the supplemental materials, FIG.s S1-S7. EA functions were estimated by considering a range of input values and equating the magnitude of the response over opposite vertical excitations ($\delta z, -\delta z$); ie., solving the expression $w'(x_{EA}/L, \delta z, V_{dc}, k_x, \delta x_+) = w'(x_{EA}/L, -\delta z, V_{dc}, k_x, \delta x_-)$ for $x_{EA}$. In this expression, $\delta x_+(\delta x_-)$ is the longitudinal strain associated $\delta z$ ($-\delta z$) for a particular model as detailed in Table II and FIG. 14. For simplicity, the bias was fixed at $V_{DC} = 3V$ for the imaging step.

Also plotted in FIG. 15 are experimentally observed EA OBD shape factors, normalized by the simultaneously acquired IDS images. They are listed in Table III and the associated image datasets are available in the supplemental material. Comparison of the plotted models and experimental data in FIG. 13 show that the closest match is the slip-stick, $\delta x = \beta|\delta z|$, with $\beta < 0$ (green, dash-solid curve). Additional details for this case and the other EA cases are discussed in Table 1. Some of the E-B ($\delta z$, $-\delta z$) solutions used to estimate this curve are shown in FIG. S4.

| Detector | Force Curve $k_x$ | Force Curve $\delta x$ | Imaging $k_x$ | Imaging $\delta x$ Figure | Legend | Origin plots | Label | EA Position ($x_{EA}/L$) | EA Shape factor ($S^*$) | Figure |
|---|---|---|---|---|---|---|---|---|---|---|
| | $10^4$ | 0 | $10^4$ | $\alpha\delta z; \alpha < 0$ [a] | - - - | S1(a, b) | Stick-stick | $x_{EA}/L \approx 0.585$ | $-15 \leq S^* \leq 15$* | |
| | $10^4$ | 0 | $10^4$ | $\beta\|\delta z\|; , \beta > 0$ [b] | ······ | S2(a, b) | Stick-stick | $0 \leq x_{EA}/L \leq 1$ | $S^* \approx 1$ | 8 |
| | 0 | 0 | $10^4$ | $\alpha\delta z; \alpha > 0$ [a] | - - - | S3(a, b) | Slip-stick | $x_{EA}/L \approx 0.515$ | $-15 \leq S^* \leq 15$* | |
| | 0 | 0 | $10^4$ | $\beta\|\delta z\|; \beta < 0$ [b] | -··-·· | S4(a, b) | Slip-stick | $0 \leq x_{EA}/L \leq 1$ | $0 \leq S^* \leq 2$** | 8 & 9 |
| OBD | $10^4$ | 0 | 0 | 0 [c] | ⊙ | S5(a, b) | Stick-slip | $x_{EA}/L \approx 0.515$ | $S^* \approx 0.851$ | 8 |
| | 0 | 0 | 0 | 0 [c] | ⊠ | S6(a, b) | Slip-slip | $x_{EA}/L \approx 0.585$ | $S^* \approx 1$ | |
| | 0 | 0 | $10^4$ | $\alpha\delta z + \beta\|\delta z\|; \alpha < 0, \beta = 0.05$ [d] | — | S7(a) | Slip-stick | $0 \leq x_{EA}/L \leq 1$ | $0.5 \leq S^* \leq 1.5$** | 9 |
| | 0 | 0 | $10^4$ | $\alpha\delta z + \beta\|\delta z\|; \alpha < 0, \beta = -0.1$ [(e] | - - - - | S7(b) | Slip-stick | $0 \leq x_{EA}/L \leq 1$ | $-0.5 \leq S^* \leq 3$ | |
| IDS | All | All | All | All | ⊞ | S1-S6(c, d) | All | $x_{EA}/L \approx 1$ | $S \approx 1$ | 8 & 9 |

Table II. The theoretical E-B calculation parameters used to estimate equal-amplitude (EA) conditions shown in FIG.s 14 and 15. For the oppositely poled vertical domains in PPLN, FIG. 14(a) – (c) show symmetry cases explored for the δx components. The six cases for OBD EA conditions are plotted in FIG. 15, appearing as colored curves or markers as denoted in the Legend column. Examples of the response curves used to estimate the EA conditions are in the supplemental materials S1-S7. Note that there are numerous EA conditions that depend on the balance of bias and in-plane contributions for OBD detection, while for the IDS (the bottom row), the EA condition is always at $x_{IDS}/\mathrm{L} \approx 1$ and S ≈ 1.

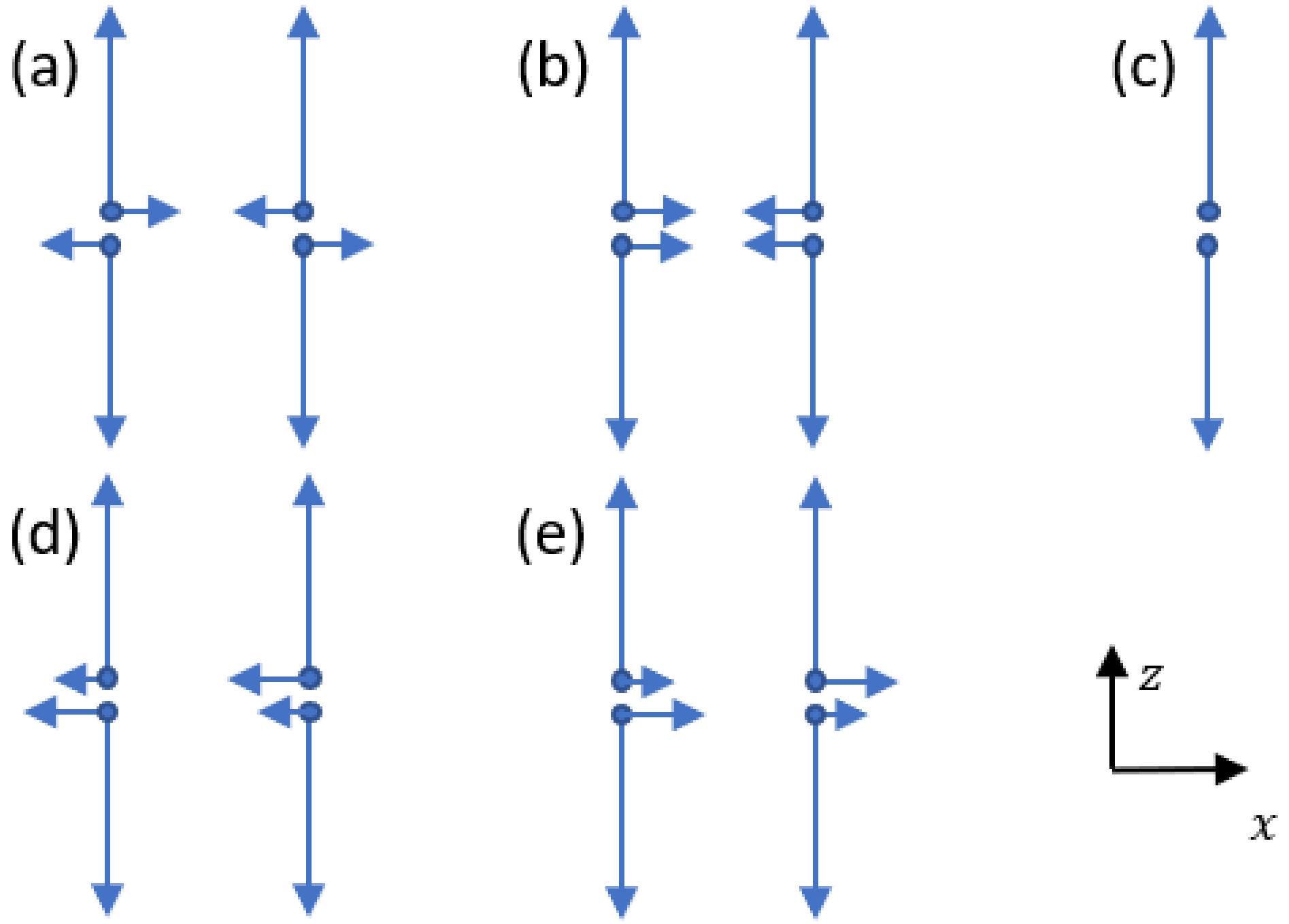

FIG. 14. Domain symmetries explored in the EAM models plotted in FIG.s 15 and 16. (a)-(c) are permutations used to develop the EA curves shown in FIG. 15. (d) and (e) are a mix of permutations (a) and(b) that are used to develop the curves in FIG. 16 that better explain the range of observed EA conditions in the experiments. These mixed model results appear as solid and dashed green curves in FIG. 16.

| Legend* | Source |
|---|---|
| ○ | EAM results, Figures 1 and 2 (GG01-07 in Suppl) |
| △ | MEFM, Figure 4(c), -10V |
| ◊ | RM, Figure 5, Trial 5 |
| □ | Scan angle test, Figure 3 (90° and 270° scans) |
| ⋇ | BB01 – in Suppl |
| ▽ | BB02 – in Suppl |
| ◁ | BB03 – in Suppl |
| ⋈ | PPLNAutoTImages – in Suppl |
| ◇ | PPLNAutoFImages – in Suppl |

Table III. Experimental Equal amplitude (EA) measurement summary and legend for FIG.s 15 and 16. The Source column lists the associated datasets.

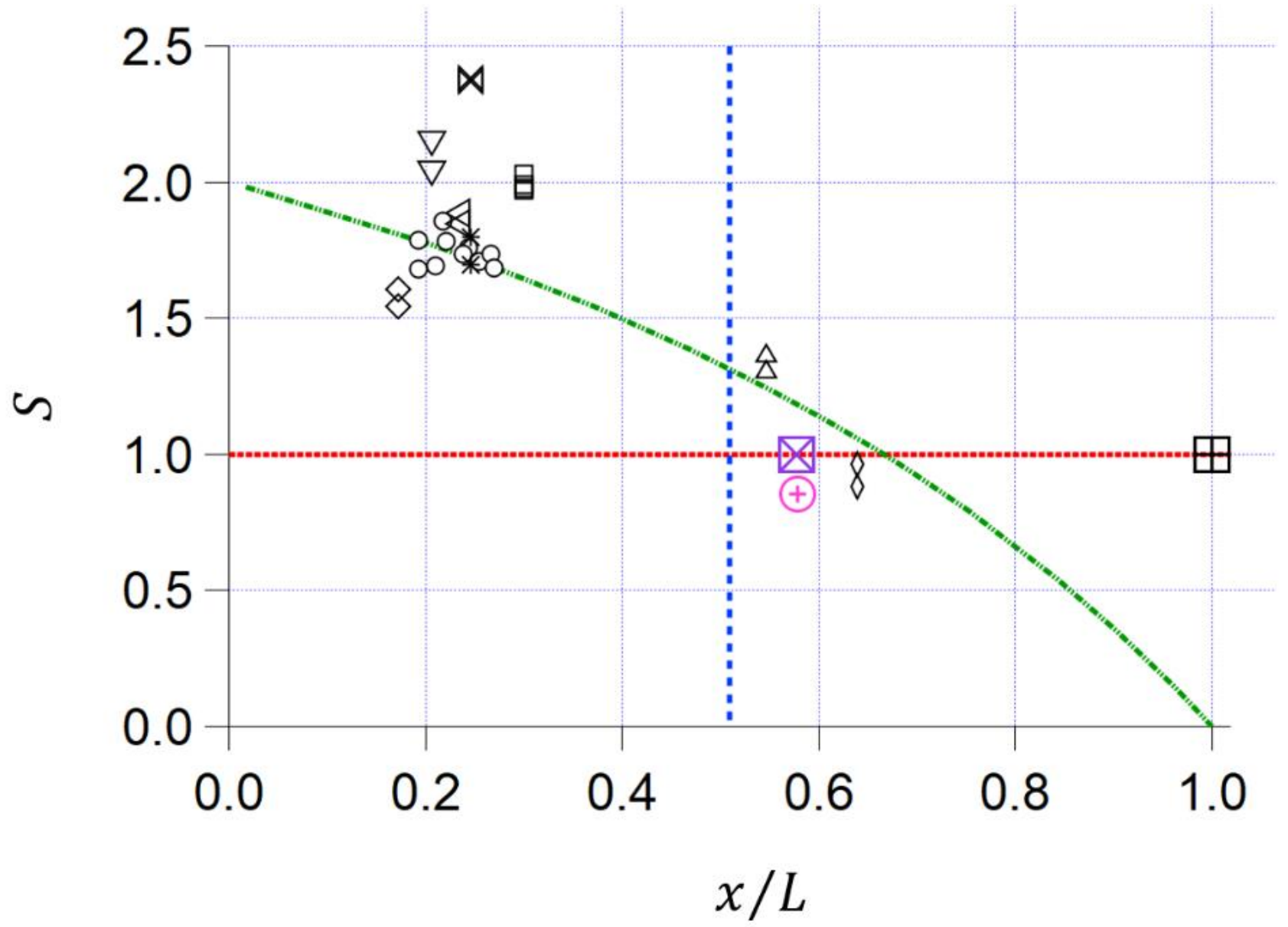

FIG. 15 shows the EA (equal amplitude) condition Shape factors versus spot position for five permutations of force curve and imaging conditions (sticking versus slipping) shown in FIG. 14. The model parameters are listed in more detail in Table II and the longitudinal symmetries are illustrated in FIG.s 14 (a)-(c). In the case of the IDS, there is a unique EA position at $x_{IDS} \approx 1$ and $S_{IDS}(x_{IDS}) \approx 1$. For the OBD, there are numerous, boundary condition-dependent EA locations, denoted as either colored lines or markers. The model that most closely reproduces the experimental EA data is given by the dashed green curve. This model is a slippery ($\mathrm{F_x} = 0$) force curve, sticky ($\mathrm{F_x} \neq 0$) imaging model with a fixed $\mathrm{V_{DC}}$ bias and a fixed, $\delta\mathrm{x} < 0$ value, independent of the sign of $\delta$z. Experimental values are plotted as black, open geometric markers.

The open geometric markers in FIG. 15 represent the OBD measurements listed in Table III. To convert from the amplitude to a shape factor, we take the IDS measurement at $x_{IDS}/L = 1$ as the ground truth. Explicitly,

$$S_{OBD} = A_{OBD}(x_{EA}/L)/A_{IDS}(x_{IDS}/L = 1), \tag{7}$$

where $x_{EA}$ is the OBD EA spot location, $A_{OBD}$ is the force curve calibrated OBD amplitude and $A_{IDS}(x/L = 1)$ is the IDS amplitude measured above the tip position, serving as the reference measurement.

Of the models in FIG. 15, the best match to the experimental results still varies substantially from the experimental measurements. FIG. 16 below shows that the majority of the experimental measurements can be explained by invoking a mixture of the components in FIG. 14 (a) and (b). Specifically, with $\delta x = \alpha\delta z + \beta|\delta z|$, as illustrated in FIG.s 14 (d) and (e), there is a rotation of the EA curve, roughly centered on the point $x_{OBD}/L \approx 0.6$, $S_{OBD}(x_{OBD}) \approx 1$.

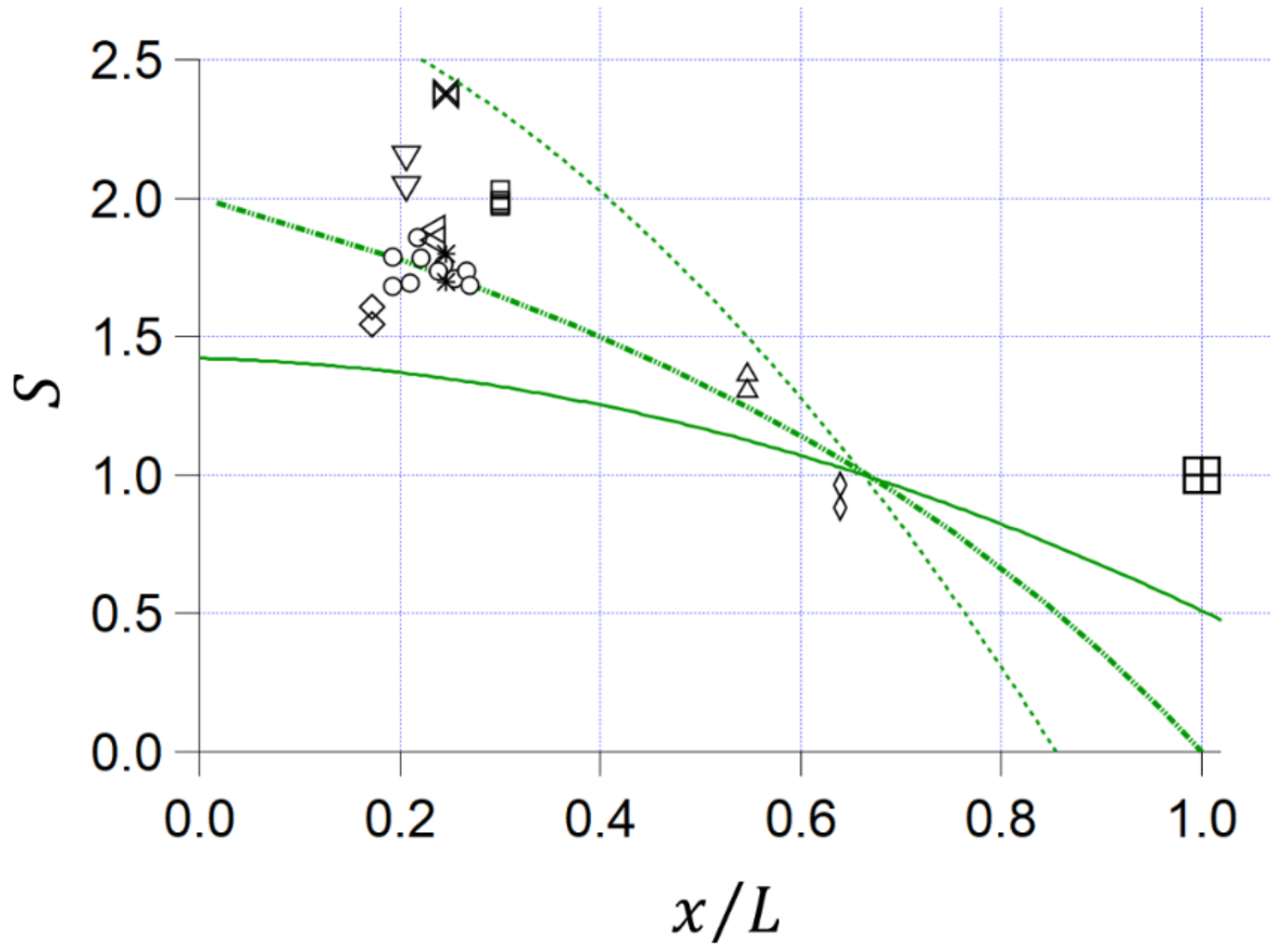

FIG. 16 shows the EA (equal amplitude) condition Shape factors versus spot position ~1. For the OBD, there are numerous, boundary condition-dependent EA locations, denoted as either colored lines or markers. The model that most closely reproduces the experimental EA data is given by the dashed green curve. This model is a slippery ($F_x = 0$) force curve, sticky ($F_x \neq 0$) imaging model with a fixed $V_{DC}$ bias and a fixed, $\delta x < 0$ value, independent of the sign of $\delta z$.

**IV. DISCUSSION**

Quantitative PFM images of PPLN as a function of spot position and load was measured using simultaneous OBD and IDS measurements to explore practical aspects of the null point approach, that is, placing the optical spot in a specific location to desensitize the measurement to BES forces acting on the body of the cantilever and in-plane (longitudinal) forces acting on the tip. Previous work has suggested that the null point locations for both the BES and longitudinal forces are $x_{OBD}/L \approx 0.6$, while for IDS, they should be at $x_{IDS}/L \approx 1$. We made many automated experiments, varying both the spot location and load, but with identical types of cantilevers and the same sample in an attempt to minimize the uncontrolled variables.

Despite using the null point approach, which improved reproducibility by placing the spot near the tip, there was still commonly to 50-100% variation between the IDS and OBD amplitude estimations.

Overall, it is important to note that PPLN is regarded as a simple material, and accurate estimates of the vertical $d_{eff}$ should be relatively easy. Other materials, including polycrystalline or disordered materials are expected to be much more challenging. If there are significant discrepancies in PPLN, it suggests that more complex samples would yield worse results. In that sense, the results of this study should be seen as a best-case scenario for those more complex materials. Most material samples will not have a convenient self-calibration between up and down domains. One might propose that we could use PPLN as a reference material where we position the OBD spot until the EAM condition is met. However, the presence of uncontrolled longitudinal (frictional) forces precludes this as a reliable method. As the data of FIG. 4 shows, the presence of in-plane forces lead to variations in the null point locations and make accurate measurements over different materials very unreliable for OBD measurements.

Unlike previous studies, this analysis considered *both* in-plane and BES forces. The basic requirement for an overall “null point” – meaning a spot position where the only surviving signal is due to the vertical PFM signal – requires the linear addition of the longitudinal shape and the BES shape to sum to zero. FIG. 13(a) shows the location of the null point in the limit that longitudinal forces vanish ($x_{OBD,F_x \to 0}/L \approx 0.58$), while FIG. 13(b) shows the null point ($x_{OBD,V_{DC} \to 0}/L \approx 0.67$). Since the zeros of those isolated excitations appear at different locations along the length of the lever for OBD, the zero location for the sum will vary, depending on the relative magnitudes of the two excitations. In fact, as our simulations show, the overall zero can occur anywhere along the length of the lever. This makes practical location of the overall null point experimentally challenging, since it may be very far away from the starting point recommended by Killgore et al. ($x_{OBD}/L \approx 0.6$). Worse, as the tip-sample interactions change due to scanning or even lateral drift, the overall null point measurement position may change, either introducing errors in the measurement or requiring spot repositioning. The same FIG.s 11(a) and 11(b) show that, for the IDS, the null points occur at the same location, ($x_{IDS,F_x \to 0}/L = x_{IDS,V_{DC} \to 0}/L \approx 1$). That means that when excitations due to longitudinal forces and excitations due to BES forces are summed, the location of the overall null point will remain at $x_{IDS}/L \approx 1$, independent of the relative magnitudes of the two contributions.

The simulations in this study were conducted in the "hard indentation limit," which we define as $k_{tip-samp} \gg k_{cant}$. As discussed in reference to FIG. 11, this improves the accuracy of the strain transferred to the cantilever since less is absorbed in the tip-sample interface. In addition, it is an assumption that greatly simplifies the modeling to the limit of a pinned tip. This limit may be experimentally out of reach for many materials, including those categorized as "strange ferroelectrics" due to the requirement of high loads and/or high stiffness cantilevers; that is a topic beyond the scope of this paper.

The summary experimental data in FIG.s 15 and 16, along with the supporting information also point out that when the EAM spot location is $x/L < 0.6$, the strain amplitude estimated with the OBD, both theoretically and experimentally is larger than that measured with IDS. This overestimation is also shown in the EAM OBD images FIG.s 8 (a) and 12(c). If we combine this observation with the additional observation that the IDS measured strain amplitude is always smaller than the limit suggested by Kalinin et al., that for PPLN, $d_{eff} \approx 12pm/V$, it is notable that all our IDS measurements to date fulfill the condition $d_{eff}^{IDS} \leq 12pm/V$. If this limit continues to hold, it implies any OBD measurements showing a value larger than that are invalid. It also implies that even OBD measurements that report $d_{eff}^{OBD} \leq 12pm/V$ may in fact be incorrect and that the vertical motion experienced by the cantilever tip may well be substantially less than the OBD number.

One of the uncontrolled variables of this and almost all AFM studies is tip shape. Although manufacturers may specify the tip sharpness, these shapes are rarely verified before the probes are used. The probe shape also changes during measurements. The tip shape and wear will control both the contact stiffness and the electric field applied to the sample surface. In the specific case of electromechanics, it is well known and demonstrated here that higher loads lead to better strain transfer to the vertical motion and therefore larger electromechanical response. This can be simply understood by the simple stiffness divider concept illustrated in FIG. 11; if the tip-sample contact is soft compared to the cantilever stiffness, much of the strain motion is absorbed by that mechanical junction and does not contribute to the observed motion of the cantilever. This can be mitigated by applying a higher load, this will increase the tip-sample contact area and stiffen $k_{tip-samp}$. However, larger loads imply faster tip wear and that may affect the conductivity of the tip in contact with the surface. Conceptually, this is similar to the stiffness divider and is also illustrated in FIG. 11. If the conductivity of the tip drops, the sample beneath the tip may experience a potential less than that being applied to the cantilever, again, reducing the strain response. This provides the PFM operator with a conundrum – high loads are required for better mechanical contacts but lead to higher friction and faster tip wear.

There are a few simple practical actions any PFM practitioner can take to evaluate for the presence of BES and longitudinal interactions mixing with the cantilever response to the vertical displacement. These simple tests are universally available on every commercial AFM known to the authors.

1. Do trace and retrace measurements match?
    a. Scan-direction dependence of measured electromechanical properties such as the amplitude and/or phase implies the existence of significant in-plane forces.

2. Does changing the scan angle change the results?
   a. Changing the scan angle changes the mix of longitudinal and lateral forces relative to the vertical and BES forces. In this work, seeing changes in the electromechanical amplitude (and phase, data now shown) as the scan angle was varied was a reliable test for revealing the existence of these forces.
   b. Note that while the IDS allowed the isolation of the vertical electromechanical response, that was not the case for the OBD measurements. We have the means to test for the presence of a mixture of forces in OBD measurements, however, accurate quantification remains elusive.

This accuracy challenge is particularly concerning in the backdrop of the dramatic growth we've seen in the past decade in automated analysis and experimentation enabled by various developments in machine-learning and artificial intelligence. PFM has been a fertile breeding ground for those sorts of studies, driven by emerging technology needs, accessible functional properties and the ability to locally measure and control functional structures and behaviors. Some recent examples include Bayesian optimization,[63] unsupervised learning control and structure of domain walls,[64] defects in PZT,[65] and a synthesis-structure-property relationship discovery tool based on a large (roughly 25k) PFM image database.[66] It is well known that models trained on these data sets will typically have microscope calibration and reproducibility limitations that stem from instrumental crosstalk, sample and probe state variations and limited data sizes – all in addition to the sample properties and functionality questions that were presumably the motivation for the measurements in the first place. To enable these exciting new capabilities on a wider scope and to avoid "garbage-in, garbage-out" scenarios, it is important for the measurements to become as accurate and reproducible as possible. Successful implementation of accurate and dependable electromechanical measurements powered by machine learning approaches holds tantalizing promise for experimental automation,[67], [68] for example, in materials combinatorics.[64]

## V. CONCLUSIONS

Following the development of the OBD detector, the vast majority of commercial AFMs moved away from interferometric detection. However, as the demand for more accurate nanoscale measurements grow, we expect this trend to reverse. The scientific method is predicated on testing conclusions in different laboratories and getting the same result. This has been an ongoing challenge for AFMs and interferometry offers a unique and powerful approach for quantifying tip-sample interactions, thereby improving this situation. Interferometry does not mitigate all of the challenges facing AFM, the AFM tip itself often remains poorly controlled or characterized. However, the combination of intrinsic calibration based on the wavelength of light, low noise coupled with a better understanding of the true motion of the cantilever represents a significant improvement in the accurate measurement of the tip-sample forces. Our IDS measurements for PPLN systematically yielded $d_{eff}^{IDS} \leq 12pm/V$. This is tantalizingly in agreement with a rigorous estimate of the effective piezoelectric coefficient predicting $d_{eff}^{PPLN} = 12pm/V$, especially when one considers that soft tip-sample contact stiffness or worn tips with poor conductivity will reduce this value.

The main thrust of this investigation was to compare OBD and IDS measurements. By using a metrological AFM that allows automated, simultaneous OBD and IDS measurements, we were able to demonstrate reproducible measurements of the vertical piezoresponse in the presence of background electrostatic and longitudinal in-plane forces with the IDS. Perhaps most informatively, simultaneous measurements allowed us to directly compare OBD and IDS measurements and uncover a key difference between IDS and OBD measurements. For OBD, because the zeros for BES and longitudinal excitations occur at different spot positions, it is extremely difficult to predict where an overall zero will appear when these shapes are added together with unknown amplitudes. Since these forces In contrast, for the IDS, since these zeros all exist immediately above the tip, we confirmed the results of previous studies that found "null points" where the effects of longitudinal or BES forces were ignored by the OBD detector, allowing correct measurements of the vertical response, we also discovered that presence of both of these effects – a typical situation in PFM – renders OBD null point methods extremely difficult to accurately quantify. One important theoretical prediction that was verified is that even when OBD measurements satisfy the equal amplitude over oppositely poled domains criteria, the OBD measurement will be incorrect by as much as 50-150%.

We also proposed some simple diagnostics, applicable for both OBD and IDS measurements that can identify the presence of significant BES and longitudinal forces in electromechanical measurements. These simple diagnostics are available on every AFM. While they do not quantify or correct for the combination of in-plane and BES forces, they do indicate whether they are present and significant in the measurements. If electromechanical measurements "test positive" for these sources of error – meaning trace and retrace do not match or the response depends on the scan angle – then extreme care must be exercised in attempting to quantitatively interpret the magnitude of the local vertical electromechanical response. Unfortunately, linear mixing of the BES, longitudinal and vertical force also implies that even relative measurements are unlikely to be quantitative. Finally, although it is beyond the scope of the present paper, even the sign of the response can be incorrect, depending on how the components combine.

As discussed in FIG. 12, accurate IDS measurement of the hard indentation limit effective piezoresponse as described by Kalinin et al. requires that the cantilever is in that limit: $k_{tip-samp} \gg k_{cant}$. In practice, this is difficult, especially in low-load, conditions with a sharp tip. However, even if the It is important to note that while this measurement shows a response $d_{eff} \leq 12pm/V$, it still does capture the physics of the tip-sample contact. The soft $k_{tip-samp}$ represents real physics affecting the vertical motion of the tip. This is fundamentally different from the case of the OBD measurements where, on top of the tip-sample there is a spot-position dependent crosstalk between the apparent vertical response and the balance of BES and longitudinal forces. To date, both in the work in this paper and in numerous other publications, the $d_{eff} \leq 12pm/V$ condition is violated. It is also interesting to speculate that, in cases where OBD measurements do fulfill the $d_{eff} \leq 12pm/V$ condition, if investigators were to simultaneously measure the actual tip displacement, they would observe amplitudes that were quite different from the OBD estimates. This was certainly the case for all the data in this work, with one exception (see FIG.s 13 and 14).

We intentionally limited this study to one type of sample – PPLN – where the domain structure was simple, the surface was smooth, clean, and flat, the electromechanical, properties were as well understood as possible and where there was a body of theoretical predictions. Even with those advantages, accurate interpretation of OBD measurements is complex and difficult. We anticipate that this problem may be practically intractable for samples where (i) we do not have that type of a-priori knowledge or (ii) samples that are more heterogeneous.

The basis for the scientific method is the ability to accurately reproduce measurements between laboratories. As the scale of the measurements gets smaller and the interaction volumes shrink, this becomes more and more difficult because noise and crosstalk become relatively more significant. These effects have undoubtedly contributed to the increasing number of reports of "strange" ferroelectrics in materials and other challenges in reproducibility. By re-introducing interferometry to AFM, we have improved the accuracy of AFM electromechanical measurements, both the IDS measurements themselves, but also as a useful tool for a better understanding of OBD measurements. Finally, the mixing of in-plane and vertical forces in AFM affects measurements well beyond electromechanics. The insights from this study have relevance for many other applications including tribology, friction and nano-mechanics and nano-rheology.

**XII. SUPPLEMENTARY MATERIAL**

See FIG.s S1-S7 show detailed Euler-Bernoulli modeling shape factor results summarized in FIG.s 15 and 16 above. FIG. S8 shows RSM angle-dependent measurements and FIG. S9 shows RSM time-dependent measurements. Finally S10-S23 show the PPLN amplitude data that are summarized in FIG.s 3, 15 and 16 above.

**ACKNOWLEDGMENTS**

RP thanks Bryan Huey, Sergei Kalinin, Denis Alikin and Andrei Kholkin for a critical reading of the manuscript and numerous helpful suggestions.